\documentclass[
]{aa}
\usepackage{indentfirst}

\usepackage{txfonts}
\usepackage{graphicx}
\usepackage{xcolor}

\usepackage{setspace}
\usepackage{hyperref}
\hypersetup{
    colorlinks,
    breaklinks=true,
    linkcolor={red},
    citecolor={blue},
    urlcolor={blue}
    }
\usepackage[all]{hypcap}

\bibpunct{(}{)}{,}{a}{}{,}

\def\Msun{\text{M}_{\odot}}

\def\Rns{{R}_\text{NS}}
\def\Mns{{M}_\text{NS}}

\newcommand{\bma}[1]{\mbox{\boldmath${#1}\/$}}

\newcommand{\Nabla}{\bma{\nabla}}
\newcommand{\E}[1]{\times10^{#1}}

\makeatletter
\renewcommand*\aa@pageof{, page \thepage{} of \pageref*{LastPage}}
\makeatother

\begin{document}
\title{Mass-loss and composition of wind ejecta in type I X-ray bursts}

\author{ Y. Herrera \inst{\ref{UPC},\ref{IEEC},\ref{ICE}}
        \and
        G. Sala \inst{\ref{UPC},\ref{IEEC}}
        \and
        J. Jos\'e  \inst{\ref{UPC},\ref{IEEC}}
        }

\institute{
    Departament de F\'isica, EEBE, Universitat Polit\`ecnica de Catalunya, c/Eduard Maristany 16, 08019 Barcelona, Spain.\label{UPC}
  \and    
    Institut d’Estudis Espacials de Catalunya, c/Gran Capit\`a 2-4, Ed. Nexus-201, 08034 Barcelona, Spain. \label{IEEC}
  \and
    Institute of Space Sciences, c/Can Magrans, 08193 Cerdanyola del Vall\`es, Barcelona, Spain.\label{ICE}
    }

\abstract { 
  X-Ray bursts (XRB) are powerful thermonuclear events on the surface of accreting neutron stars (NS), where nucleosynthesis of intermediate-mass elements occurs. 
  The high surface gravity prevents the ejection of material directly by the thermonuclear explosion. 
  However, the predicted and observed luminosities sometimes exceed Eddington's value, and some of the material may escape by means of a stellar wind. 
}{
  This work aims at determining the mass-loss and chemical composition of the material ejected through radiation-driven winds and its significance for Galactic abundances.
  It also reports on the evolution of observational quantities during the wind phase, which can help constrain the mass-radius relation in neutron stars.
}{
  A non-relativistic radiative wind model was implemented, with modern opacity tables and treatment of the critical point, and linked through a new technique to a series of XRB hydrodynamic simulations, that include over 300 isotopes. 
  This allows us to construct a quasi-stationary time evolution of the wind during the XRB.
}{
  In the models studied, the total mass ejected by the wind was about $6\E{19}\ \texttt{g}$, the average ejected mass per unit time represents $2.6\%$ of the accretion rate, with $0.1\%$ of the envelope mass ejected per burst and $\sim 90\%$ of the ejecta composed by $^{60}$Ni, $^{64}$Zn, $^{68}$Ge and $^{58}$Ni.
  The ejected material also contained a small fraction ($10^{-4}-10^{-5}$) of some light p-nuclei, but not enough to account for their Galactic abundances.
  Additionally, the observable magnitudes during the wind phase showed remarkable correlations, partly deriving from to the fact that photospheric luminosity stays close to Eddington limit. 
  Some of these correlations involve wind parameters like energy and mass outflows, that are determined by the conditions at the base of the wind envelope.
}{
  The simulations resulted in the first realistic quantification of mass-loss for each isotope synthesized in the XRB.
  The photospheric correlations found could be used to link observable magnitudes to the physics of the innermost parts of the envelope, close to its interface with the NS crust. 
  This is a promising result regarding the issue of NS radii determination. 
}


\maketitle

\section{Introduction} \label{sect:intro}

  Type I X-ray bursts (XRBs) are highly energetic and recurrent thermonuclear events occurring on the envelope of accreting neutron stars (NS) in binary systems where the secondary star is usually a main sequence star or red giant.
  Most observed XRBs have short orbital periods, in the range 0.2--15 hr.\footnote{
    Exceptions include GX 13+1 (592.8 hr), Cir X-1 (398.4 hr), and Cyg X-2 (236.2 hr).
  }
  As a result, the secondary star overfills its Roche lobe and mass transfer ensues through the inner Lagrangian point (L1) of the system. 
  The material stripped from the secondary has angular momentum, such that it forms an accretion disk around the NS.
  Viscous forces progressively remove angular momentum from the disk forcing the material to spiral in and pile up on top of the NS.
  The accreted material accumulates under mildly degenerate conditions, driving a temperature increase and the onset of nuclear reactions. 
  As a result, a thermonuclear runaway occurs, generating a massive luminosity increase as well as nucleosynthesis of intermediate-mass elements, mostly around $A = 64$ \cite[see, e.g.,][]{Woosley_2004,Fisker_2008,Jose2010}.
  The presence of such elements can, in principle, be detected  \cite[see][]{BildChangPaer2003,ChangBildWasser2005,ChangMorBildWaser2006,WeinBildScha2006} in the form of gravitationally-redshifted absorption features in the spectrum, which mostly lies in the X-ray range. 
  For further information on XRBs see \cite{StrohmBild2003, KeekZand2008, GalloMunHartChak2008,JoseBook2016}.

  A typical NS ($\Mns \sim 1.4 \ \Msun$, $\Rns \sim 10 \ \texttt{km}$) has 
  a strong surface gravity,
  allowing only a limited envelope expansion before the nuclear fuel is consumed.
  However, for some values of accretion rate, a considerable photospheric radius expansion (PRE) takes place.
  In these cases, the luminosity can approach or even exceed the Eddington limit in the outer layers of the expanded envelope, which may lead to the ejection of some material through a radiation-driven wind \cite[see][and references therein]{HSJ2020}.
  The potential impact of XRBs on the Galactic abundances is still a matter of debate.

  Additionally, the study of XRBs wind can lead to a more accurate determination of neutron stars radii. 
  Several theoretical mass-radius relations are available for NS, depending on the choice of equation of state when modeling their inner structure 
  \cite[see][]{Lattimer+Prakash-2006}. 
  Thus, accurate measurements of NS masses and radii, taken independently, are needed to test the validity of these theoretical models.
  Neutron star masses, for instance, can be determined from bolometric luminosities during XRBs with significant envelope expansion (assuming the maximum corresponds to the Eddington limit).
  The radius is more difficult to measure, and methods proposed carry a varied sensitivity to systematic errors \cite[see][]{Damen++1990,Steiner2010,Guver++2012,Sala++2012}.
  A detailed study of the NS envelope evolution during XRBs, that includes stellar wind features, can bring more light into this matter.
  
  Motivated by the aforementioned unanswered questions and by the new possibilities brought by the recent technical advancements, the interest of the scientific community in stellar winds in the context of XRBs has been renewed. 
  For instance, \cite{YuWeinberg2018} used MESA code \cite[see][]{MESA1} to perform a 
  hydrodynamic simulation of the wind after the Eddington limit is reached
  during a hydrostatic burst; and \cite{Guichandut2021} 
  studied the transition from static expanded envelopes to radiatively-driven stellar wind and discussed the applicability of steady-state models during it.
  Results from our simulations of radiative winds applied to a generic NS scenario were published in \cite{HSJ2020}.
  There, the focus was set on the exploration of parameter space, the characterization of different types of solution and their self-consistency with model hypotheses, and analysis of possible predictions related to observable variables, paving the road to the application of the wind models to a more complex XRB scenario. 
  
  In this work we apply the stellar wind model, 
  implemented through a simulation code entirely developed for the present work, to the physical conditions of type I XRBs. 
  We proceed towards this goal by matching numerical solutions of the stellar wind model, to a set of XRB hydrodynamic models
  \cite[]{Jose2010}.
  This way, we are able to obtain physical predictions about the evolution of the entire neutron star atmosphere, including both the nuclear burning shells and the expanded wind envelope. 
  On one hand, this allows us to study the evolution of observable magnitudes in more realistic conditions, and their relation to parameters such as NS mass and radius, potentially leading to the development of better measuring techniques. 
  And on the other hand, we can obtain the composition of the envelope layers blown by the wind.
  This allows us to study the possible contribution of XRBs to Galactic abundances.
  
  This work is organized as follows.
  In Sect. \ref{chap: XRB_Wind}, a summarized description of both stellar wind and 
  XRB hydrodynamic models is followed by an account of the methods employed to match 
  them.
  An analysis of the results obtained in our best matching model, including the wind-related mass-loss and its contribution of key isotope species to galactic abundances, as well as the impact on observable features, is shown in Sect. \ref{sect: XRB-wind results}.
  Finally, the significance of our results and our main conclusions are summarized in Sect. \ref{sect:discuss}.

\section{Models, input physics, and methods} 
\label{chap: XRB_Wind}
  
  Our objective is to determine if the physical conditions given by the XRB hydrodynamic models are suitable for a stellar wind to appear at some point during the evolution of the burst.
  In other words, we want to find if there is a compatible match in physical variables between the possible wind profiles, obtained as solutions of the model developed in \cite{HSJ2020}, and the boundary values taken as a result from the XRB hydrodynamic models developed by \cite{Jose2010} (and unpublished data).
  These extra boundary conditions will constrain the free parameters in the wind model (see Sect. \ref{sect: wind model}), and could determine a unique wind profile at each time, in case a solution exists. 
  We are going to work under the assumption that the chemical composition of the whole wind envelope will be determined by the XRB hydrodynamic models at the matching point, for each time step.
  That is, we re-compute wind profiles adjusting all free parameters (including mean molecular mass) to match those resulting from the hydrodynamic model at each time-step (see Sect. \ref{sect: XRB-wind match}).
  If successful, the ultimate goal is to study the consequent mass-loss and predicted observable magnitudes.

  An important aspect to take into consideration is the fact that wind and XRB models differ in some key features and basic hypotheses.
  For instance, the wind model describes a stationary regime, while XRB models are time dependent.
  However, if the evolution of XRB conditions is slow enough, under some criteria, a matching sequence of steady wind solutions could still be considered a good quasi-stationary approximation. 
  Convective energy transport, degenerate matter contributions to the equation of state, and energy generation by nuclear reactions are also only contemplated in the XRB hydrodynamic models, since they were not necessary for wind models. 
  These issues will be taken into consideration when assessing the validity of the results. 

  We will first summarize the relevant features of both stellar wind (Sect. \ref{sect: wind model}) and XRB hydrodynamic models (Sect. \ref{sect: XRB models}). 
  The models matching technique is described in section \ref{sect: XRB-wind match}.

\subsection{Stellar wind model} \label{sect: wind model}

  The simulations reported in this work rely on a stationary, non-relativistic wind model with spherical symmetry.
  The radiation-driven wind, treated as a fully-ionized perfect gas, is assumed to be optically thick and in local thermodynamic equilibrium (LTE) with radiation. 
  In this framework, the basic hydrodynamic equations for mass (\ref{Eq basic mass}), energy (\ref{Eq basic energy}), momentum conservation (\ref{Eq basic momentum}), and radiative energy transport (\ref{Eq basic transport}) become:
\begin{align}
  \label{Eq basic mass}  
  \dot M &= 4 \pi r^{2} \rho v 
  \\
  \label{Eq basic energy}  
  \dot E &= \dot M \left( \frac{v^{2}}{2} - \frac{G M }{r} + h \right) +L_{R} 
  \\
  \label{Eq basic momentum} 
  0 &= v \frac{\text{d}v}{\text{d}r} + \frac{G M }{r^{2}} + \frac{1}{\rho} \frac{\text{d}P}{\text{d}r} 
  \\
  \label{Eq basic transport}  
  \dfrac{\text{d}T}{\text{d}r} &= -\dfrac{3 \, \kappa \rho L_{R}}{16 \pi c a r^{2} T^{3}},
\end{align}
  where the variables $r,T,v,\rho, L_\texttt{R} $ stand for the radial coordinate, temperature, wind velocity, gas density, and radiative luminosity, respectively. 
  The total pressure $P$, and specific enthalpy $h$, include the contributions of gas and radiation, in the form:
\begin{align}
  P{(\rho,T)} &= \frac{\rho k T}{\mu m_A} +\frac{a T^{4}}{3} 
  & \text{and}
  &&
  h{(\rho,T)} &= \frac{5}{2} \frac{k T}{\mu m_A} + \frac{4}{3} \frac{a T^{4}}{\rho}.
  \label{eq: Pressure and enthalpy}
\end{align}

  In this work, the opacity $\kappa$ was taken from 
  \textit{OPAL} \cite[]{OPAL1996} and \textit{The Opacity Project} \cite[]{TOP1994},
  and it is a function of temperature, density, and composition of the gas only.\footnote{
    The latter were especially useful since they allow changes in gas composition more conveniently, and higher values of metallicity. However, for such values they are limited to $\log T < 8$. For higher temperatures, opacity was extrapolated using a formula introduced by \cite{Paczynski1983}.
  }

  The total mass and energy outflows $(\dot M, \dot E)$ arise as integration constants from the mass and energy conservation laws and can be determined by imposing extra  conditions at the base of the wind. 
  Therefore, $\dot M$ and $\dot E $ can be considered free model parameters. 
  During an XRB these will vary in time but, in a quasi-stationary approach, are assumed to be approximately constant across the wind for a fixed time.
  The mass of the neutron star core, $M$, is another model parameter, considered fixed in this work, and constitutes the only relevant source of gravity (the contribution of the envelope mass can be neglected). 
  Lastly the mean molecular mass $\mu$ is a function of the mass fractions $X_i$ of the different species present in the envelope.
  The rest of the symbols have their usual meaning: $G$ for gravitational constant, $c$ for speed of light, $k$ for Boltzmann constant, $a = \frac{4\sigma}{c}$ for radiation energy density constant ($\sigma$ is the Stefan-Boltzmann constant) and $m_A$ for atomic mass unit.

  Two boundary conditions are required to solve the system of two first-order differential equations \ref{Eq basic momentum} and \ref{Eq basic transport}. 
  These are normally given at the photosphere (usual for any stellar envelope, but with a slight difference for winds), 
  and a critical point (for stellar wind solutions with non-zero velocity at large radii). 

  As stated before, an extra set of boundary conditions can be set to fix the value of the free model parameters $\dot M$ and $\dot E$, 
  adjusting the resulting solution to describe different physical scenarios. 
  In \cite{HSJ2020}, we explored a subset of possible wind solutions that are compatible with a very general boundary condition expected at the base of the envelope 
  for neutron stars with radius lying within a wide possible range ($7-20 \texttt{ km}$).\footnote{
    There, we chose a wind base corresponding to the point where the wind transitions from gas pressure-driven to radiation-driven (i.e. $\nabla P_R \geq \nabla P_g $).
  }
  We found that, for a particular neutron star radius of $ 13\texttt{ km}$, a sequence of solutions aligned in the parameters space, with correlated energy and mass outflows. 
  Although these values were within the expected ranges for an XRB, no specification was made on whether these solutions corresponded to actual XRB conditions or how do they evolve in time.
  Here we want to connect the stellar wind models to a more realistic burst scenario as described through the aforementioned XRB hydrodynamic models.

\subsection{XRB hydrodynamic simulations} \label{sect: XRB models}

  The XRB hydrodynamic models analyzed in this work 
  couple a nuclear reaction network, that includes 324-isotopes linked by 1392 nuclear interactions, into a modified version of the \textit{SHIVA} hydrodynamic evolution code  \cite[see][]{Jose-Hernanz-1998,JoseBook2016}. 
  The main features and hypotheses of this code are: spherical symmetry, Newtonian gravity, and energy transport by radiative diffusion and convection.\footnote{
    General relativity corrections can be determined a posteriori and rely on the factor $(1+z)$, where $z$ is the gravitational redshift \cite[see discussion section in][]{Jose2010}.
  } 
  Equation of state includes contributions from the electron gas (with different degrees of degeneracy), the ion plasma, and radiation; Coulomb corrections to the electronic pressure are also taken into account. 
  Aside from the energy generated by nuclear reactions, models took into account neutrino energy losses as well.
 
  In the present work we will focus on a model computed with the following physical input parameters: $\Mns =  1.4 \,\Msun$, $\Rns=13.1\text{ km}$, accreted material of solar composition and an accretion rate of $ 1.75 \E{-9} \,\Msun/\text{yr} $. 
  All bursts from this model show similar resulting features: burst recurrence times ranging in about $5-6.5 \text{ hr} $, burst durations of $55-75 \text{ s} $, peak luminosities within $\sim [1-2] \E{5} \text{L}_\odot$, and peak temperatures in $[1-1.25] \E{9} \text{ K} $. 
  We will show the study on one burst corresponding to a 200-shell envelope simulation, which we will call XRB-A 
  \cite[Model 2 in][and unpublished data]{Jose2010}.\footnote{
    A study on lower resolution bursts \cite[e.g., Model 1 in][]{Jose2010} yielded similar resulting characteristics, but with lower matching accuracy and lower overall mass ejected by the wind, with varying composition. In both cases, the unpublished data mentioned includes detailed chemical composition and additional physical properties of each numerical shell.
    }

    It is worth remarking that, even though \cite{Jose2010} report that no matter is ejected in their simulations, this refers to \textit{explosive} ejection (i.e., material reaching escape velocity). 
    They also state that the ejection of a tiny fraction by radiation-driven wind (i.e., when luminosity exceeds Eddington limit) is still possible.

\subsection{Models matching method} \label{sect: XRB-wind match}

  Ideally, at a given matching point, all relevant physical variables should be continuous functions. 
  However, given that both models have different hypotheses corresponding to different physical regimes the ideal match may not always be sensible or possible. 
  For instance, we can not expect a realistic match whenever degenerate gas pressure or convection are relevant terms in the equations of XRB hydrodynamic models, given that they are not present in the wind models. 
  Nevertheless, whenever the conditions are such that these terms can be neglected, we could expect to find a valid match.
 
  Another aspect to take into consideration is whether energy or mass outflows, $\dot E$ and $\dot M$, resulting from the conditions in the XRB hydrodynamic model, lie in the region of the wind parameter space where solutions can be found and are consistent with the wind model hypotheses. 
  In this regard, the study of the wind parameter space discussed in 
  \cite{HSJ2020} resulted of great significance. 
  It is worth keeping in mind, though, that the location and shape of this region also depends on the abundances of elements present in the envelope, especially hydrogen, and on the mass of the neutron star. 
  The latter is fixed for the XRB hydrodynamic models considered in this work, but the former changes dynamically within the evolution of each XRB and across its shells.
  The wind model free parameters, $\dot E$ and $\dot M$, can be uniquely determined by imposing two additional boundary conditions, for instance the temperature and density, at a particular matching radius.
  The energy outflow $\dot E$ can be equated to the total luminosity reported in the XRB models, which varies with radius, since the envelope in the XRB hydrodynamic models extends over a few meters above the neutron star core.
  In practice, this translates into finding which shell has a better match.
 
  The general idea for the matching procedure employed is as follows. 
  For a given XRB hydrodynamic simulation, we explore the shell/time-step grid, focusing first on every point with an energy outflow $\dot E$ that lies within the range of acceptable wind solutions in the parameter space.
  Since this range depends on the current element abundances and its exact extension is not known a priori, a typically safe and reduced region is chosen in terms of the ratio $\dot E /L_X$, for a first of several grid sweeps.
  In each sweep, every selected grid point is tested for a possible match by trying different wind profiles with varying mass outflows $\dot M$, while the rest of the wind parameters ($\dot E, \mu(X_i)$) is fixed to those given by the current XRB simulation grid point, until we find the one that matches one of the boundary conditions at the radius given by the current grid point. 
  For this step (let us call it ``first'' match) we chose a match in density.
  That is, we use a suitable root-finding method to find zeros of the following $\delta_\rho$ function that parametrizes the match relative error as a function of the mass outflow:
  \footnote{ It is worth mentioning here that, since the valid domain for $\dot M$ is not known a priori, derivative-based root-finding methods like the commonly used Newton-Raphson are not well suited for the task since they not only may wander outside any initial guess of a valid domain, but they also require a numerical approximation of the derivative of the $\delta_\rho$ function, which demands calculating additional wind model solutions at each step. 
  A custom root-finding algorithm, based on the safer (but often slower) bisection method, was developed for dealing with this problem.
  }
\begin{align}
  \label{eq: W2B match error 1}
  \delta_\rho(\dot M) = \frac{\rho_W(\dot M)}{\rho_B} - 1
\end{align}
  where the subindices $W$ and $B$ stand for ``wind'' and ``burst'' (i.e., from the hydrodynamic simulation) values at the matching radius, respectively. 
  Grid points where this match relative error is smaller than a desired threshold (e.g. $\delta_\rho < 0.01$) are considered successful first matches or ``hits''.

  After the first sweep is completed, the exploration is sequentially extended to previously unexplored grid points that lie adjacent to those for which a ``hit'' was found.
  The whole process is repeated until no further hits are found. 
  The aim is to avoid missing out possible hits outside the first guessed safe region, while at the same time preventing the exploration to waste efforts (and computing time) by unnecessarily testing points with parameters outside the wind model acceptable region.

  At each point, we also compute a ``second'' matching error for temperature, $\delta_T$, with an expression analogous to Eq. \ref{eq: W2B match error 1}, in order to construct the total matching error:
  \footnote{
    Several pairs of matching variables could be used here ($(\rho,T)$, $(P_R,P_g)$, etc), and in any order. 
    Which variable is chosen for the first match will affect the precision requirement of the root-finding employed, as well as the overall error obtained for the second match. 
    After trying several combinations, we found that the best results for the method employed, in terms of number of matches, speed and matching errors, is obtained with the ordered pair $(\rho,T)$.
  }
\begin{align}
  \label{eq: W2B match error total}
  \delta = \sqrt{\delta_\rho^2 + \delta_T^2}
  .
\end{align}
  Grid points where this match error is lower than a desired threshold
  are then considered to be successful XRB-wind matches.
  Matching points are tested for hypothesis consistency in both XRB hydrodynamic and wind models.

\section{Results}
\label{sect: XRB-wind results}

  The search for XRB-wind model matching points was successful for some of the XRB hydrodynamic simulations studied. 
  This means that these XRB hydrodynamic models give way to a brief stellar wind and the consequent mass-loss.

\subsection{Wind-burst matching solutions}
\label{sect: XRB-wind match analysis}

\begin{figure}
  \centering 
  \includegraphics[keepaspectratio=true,width=1.0\columnwidth,clip=true,trim=0pt  0pt 0pt 40pt]{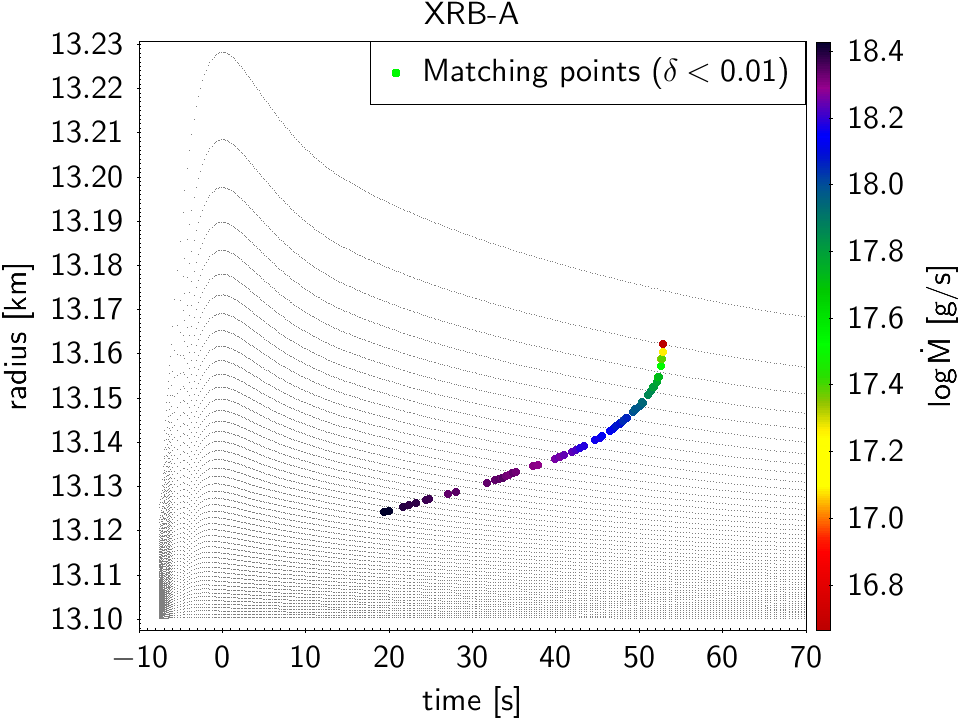}
  \caption{ 
    Wind matching points during XRB evolution (model XRB-A).
    Grey dotted lines show the radial expansion of each XRB model shell as a function of time. 
    Points for which matching wind solutions were found ($\delta < 0.01$, see text) are marked with colored circles. 
    Color indicates mass outflow $\dot M$ of the matching wind solution.
  }
  \label{fig: XRB-wind match} 
 \end{figure}

  The wind-burst matching analysis for the hydrodynamic burst model, XRB-A, resulted in the sequence of matching points shown in Fig. \ref{fig: XRB-wind match}. 
  It shows a very well defined progression from the inner (hotter and denser) layers to the outer (cooler and less dense) ones and allows us to reconstruct a continuous evolution of all physical variables (see Sect. \ref{sect: XRB-wind results mass-loss} and \ref{sect: XRB-wind results observables}).
  The matching error (Eq. \ref{eq: W2B match error total}) for these points was significantly low: $\delta < 0.01$. 
  The resulting mass outflow, $\dot M$, color-coded in Fig. \ref{fig: XRB-wind match}, evolves from higher values at the inner matching layers, diminishing as the matching point moves outwards, until no further match was found.
  Matching points for deeper layers (and earlier time) than the ones shown can be found, with the same method.
  However, in that direction, the XRB models enter a region of temperature and density where the effects from electron degeneracy and energy generation from nuclear reactions become significant. 
  Convective energy transport also becomes relevant in deeper layers.
  Wind model does not contemplate these effects, therefore possible matches towards deeper layers were not explored further.

\begin{figure}
  \centering 
  \includegraphics[keepaspectratio=true,width=1.0\columnwidth,clip=true,trim=0pt  0pt 0pt 40pt]{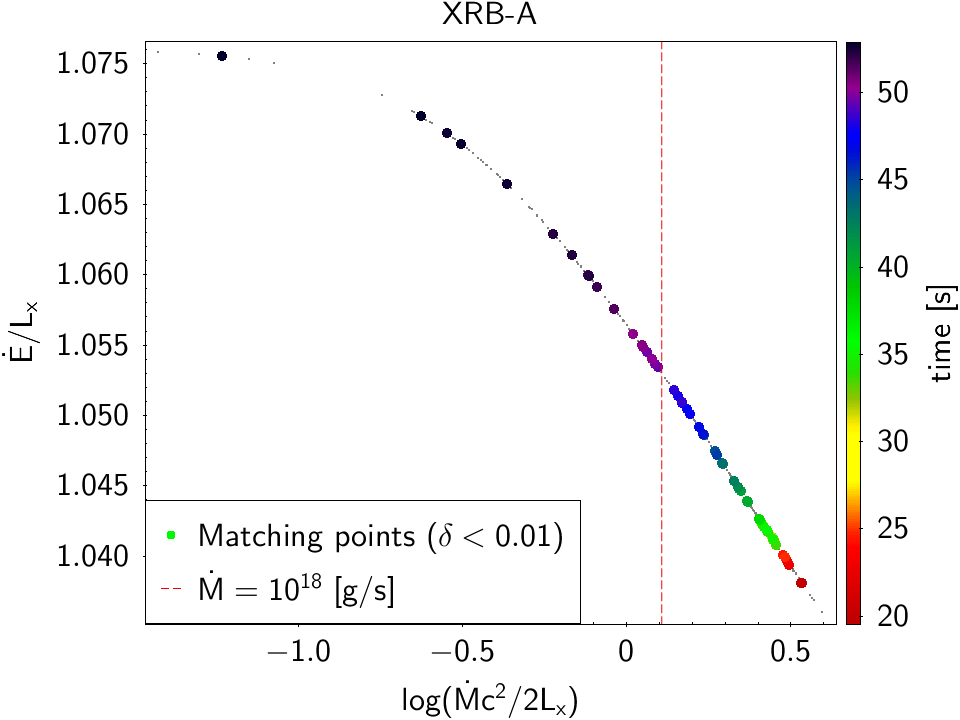} 
  \caption{
    Evolution of Wind-Burst matching points in parameter space (for model XRB-A). Colored dots represent matching points with relative error $\delta < 0.01$.
    Color scale corresponds to time since burst peak expansion.
    Both energy outflow, $\dot E$, and mass outflow, $\dot M$, are normalized in terms of Eddington luminosity in the electron scattering case, $L_X$ (see text).
    A reference value for $\dot M = 10^{18} \texttt{ g/s}$ is indicated by a vertical dashed red line.
  }
  \label{fig: XRB-Wind WiMP} 
\end{figure}

  The distribution of matching solutions in the wind parameter space is shown in Fig. \ref{fig: XRB-Wind WiMP}.
  At a first glance, the sequence of matching points display a similar distribution in parameter space to the one found in our previous work
  \cite[see, e.g., Fig. 4 in][]{HSJ2020},
  but the overall energy outflow is slightly higher here.
  Wind parameters are now normalized in terms of a non-constant Eddington luminosity corresponding to electron-scattering, $L_X$, which depends on local hydrogen mass fraction, $X$, as:
\begin{align}
  L_X = \frac{4\pi cGM}{\kappa_o (1+X)} \simeq \frac{ 3.52 \times 10 ^{38}}{ (1+X) }  \texttt{ erg s}^{-1} 
\end{align}
  with $\kappa_o = 0.2 \texttt{ cm}^2 \texttt{g}^{-1}$, as opposed to the aforementioned previous work where $X=0$ and normalization was therefore constant. 
  Otherwise, each of the matching points would deviate slightly from the observed aligned disposition, according to their corresponding hydrogen abundance.
  We can also follow the time evolution of the parameters now, given by the color scale.
  Wind total energy outflow evolves from lower to higher values, as opposed to mass outflow. 
  The matching point time progression in the parameter space is slower for earlier times and speeds up considerably in the final few seconds.

\begin{figure*}
  \centering 
  \includegraphics[keepaspectratio=true,width=0.45\textwidth,clip=true,trim=0pt  0pt 0pt 35pt]{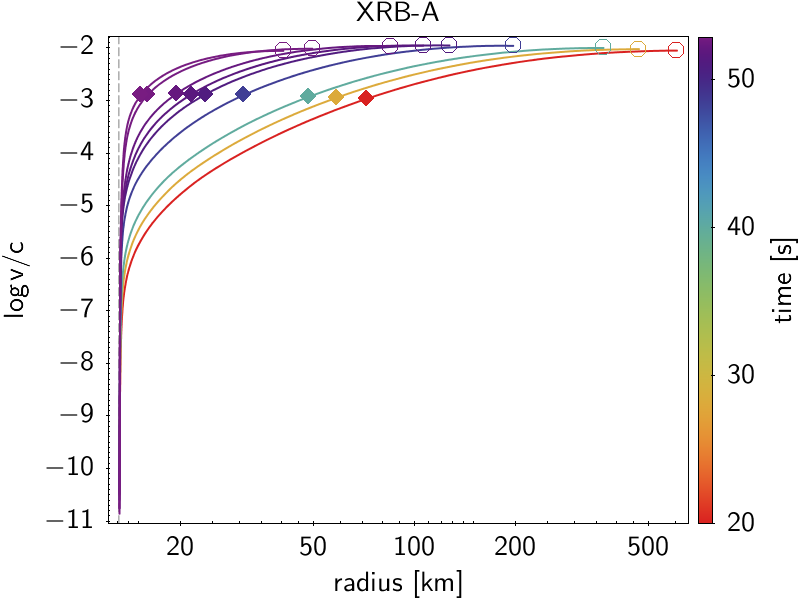}  \quad
  \includegraphics[keepaspectratio=true,width=0.45\textwidth,clip=true,trim=0pt  0pt 0pt 35pt]{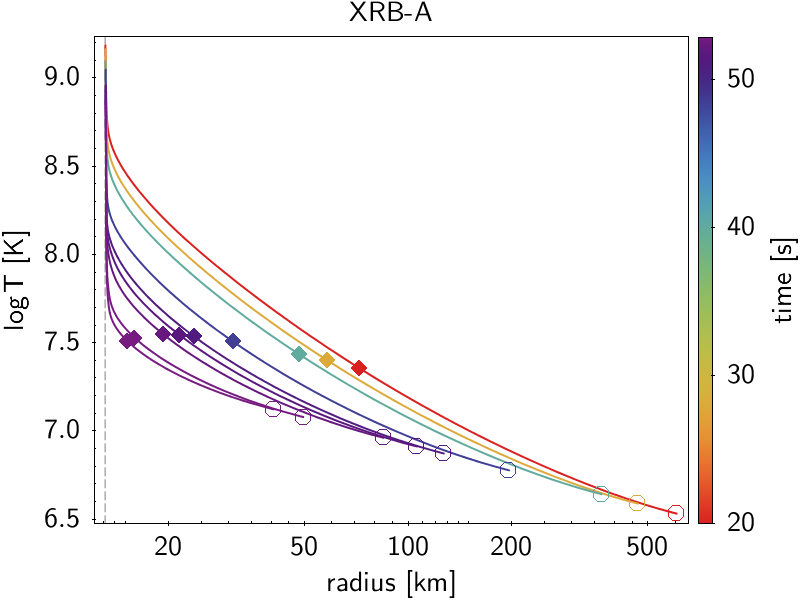} 
  \linebreak \linebreak
  \includegraphics[keepaspectratio=true,width=0.45\textwidth,clip=true,trim=0pt  0pt 0pt 35pt]{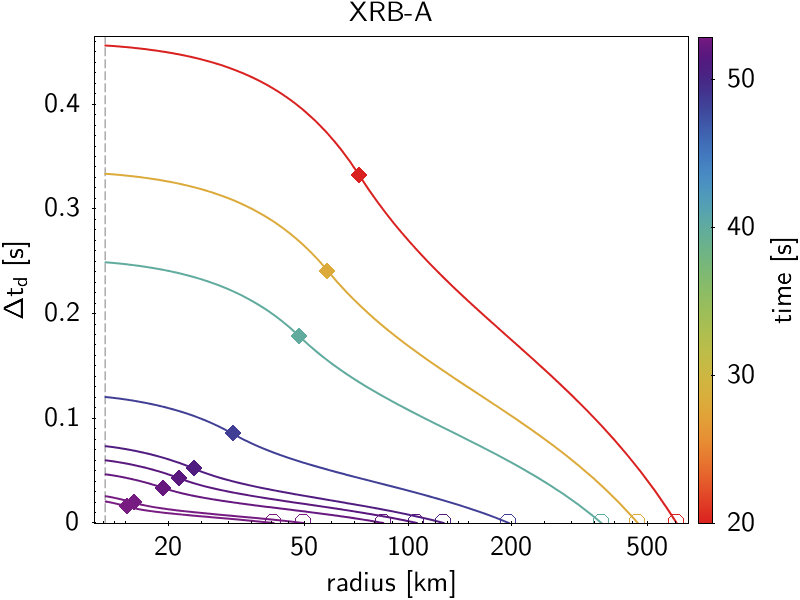}  \quad
  \includegraphics[keepaspectratio=true,width=0.45\textwidth,clip=true,trim=0pt  0pt 0pt 35pt]{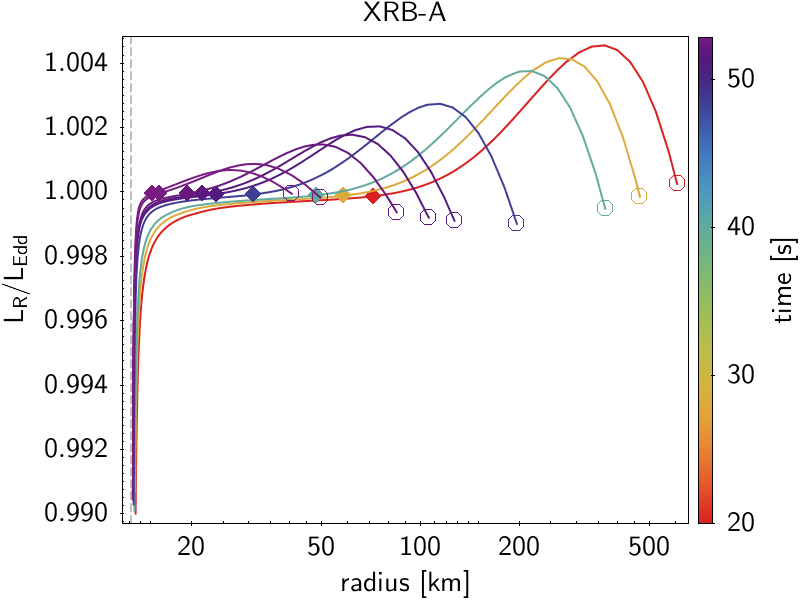}
  \caption{
    Wind profiles obtained for model XRB-A. Time evolution is indicated by line color.
    Locations of the critical sonic point ($\blacklozenge$) and photosphere ($\circ$) are indicated in each curve.
    Panels in reading order are: velocity, temperature, characteristic time, and luminosity ratio $\Gamma$ 
    (i.e. $L_R$ in terms of local Eddington luminosity, $L_\text{Edd}$),
    all plotted as a function of radius. 
    Neutron star radius is $13.1 \ \texttt{km}$ (vertical dashed line). 
  }
  \label{fig: XRB-Wind Profiles} 
 \end{figure*}

  A selection of radial profiles, for some of the matching wind solutions, is shown in Fig. \ref{fig: XRB-Wind Profiles}.
s  The profiles show similar features to the ones reported in \cite{HSJ2020}.
  The main difference is that the overall radial extension of the wind profiles shown here is smaller, which is characteristic of wind solutions with higher energy outflows.
  The radial extension of the wind envelope shrinks with time,  as it is made evident by the position of the critical (sonic) point and photosphere, indicated in each profile. 
  This is expected for the final part of the wind evolution, where the expanded wind envelope should recede and finally meet the original boundaries reported by the XRB hydrodynamic models, as the wind stops.

  Values of wind velocity at the photosphere and critical point do not vary much ($v_\texttt{ph} \sim 10^{-2}c$, $v_\texttt{cr} \sim 10^{-3}c$) for most of the duration of the wind, in a similar way to the solutions reported in \cite{HSJ2020}.
  Temperature values at these points are also similar ($\log T_\texttt{ph}(\texttt{K}) \sim 6.5-7$, $\log T_\texttt{cr}(\texttt{K}) \sim 7.5$), minding the fact that now the profiles extend only to about $500\texttt{ km}$, corresponding to $\log r(\texttt{km}) \simeq 2.7$ in Fig 1 from \cite{HSJ2020}.
  The radiative luminosity ratio profiles, $\Gamma = L_R/ L_{Edd}$, also exhibit a similar behavior to the ones reported in 
  \cite{HSJ2020}. 
  They all show a sharp rise close to the envelope base, quickly stabilizing outwards around a value of $\Gamma \sim 1$, with tiny variations of about $10^{-3}$.
  This is also a sign of a quick change from gas pressure-driven to radiation-driven wind.

  In \cite{HSJ2020} we showed a characteristic time plot following the prescription used by \cite{QuinnPacz1985}. 
  However, according to \cite{Guichandut2021}, this flow time is not a good measure of the feasibility of the quasi-stationary approach. 
  They stated the problem is that it is highly dominated by the very slow wind velocity at the wind base. 
  Instead, they proposed a combined characteristic time using the much higher sound speed for the subsonic part, and the regular wind speed for the supersonic part, giving a much smaller characteristic time than the typical wind duration.
  That is, in general, for any starting radius $R$ below the photosphere, this new dynamic characteristic time would be:
  \begin{align}
  \label{eq: characteristic time}
      \Delta t_d (R) 
      = \int_{R}^{r_\text{ph}} \frac{\text{d}r}{\max \left\{ c_s(r),v(r)\right \} }
  \end{align}
  where 
  $c_s$ is the sound speed, and the other symbols have the usual meaning.
  The total dynamic time of a wind solution is then obtained by evaluating at the wind base radius, i.e. $R=r_\text{wb}$.
  The reasoning behind this choice can be understood as follows.
  A small change in physical conditions at the base can be thought of as a perturbation from the stationary regime solution. 
  This perturbation will travel at the speed of sound as long as the flow is subsonic, readjusting the profile in its way up, and will travel with the flow velocity when it is supersonic. 
  This effectively means that $\Delta t_d$ is associated to the propagation speed of information through the wind envelope.
  As long as this information travel time is small enough compared to the duration of the wind, we can consider the quasi-stationary regime as a good approximation. 
  Given the fact that $\Delta t_d<0.5 \ \texttt{s}$ in all profiles (as shown in Fig. \ref{fig: XRB-Wind Profiles}, bottom-left panel), and that the wind duration is above $30 \ \texttt{s}$, we can safely conclude that the quasi-stationary hypothesis holds. 
  A similar argument could be made to test another assumption, which is that the composition of the whole wind profile is determined by the composition of the matching point (see Sect. \ref{chap: XRB_Wind}). 
  Changes in composition have an impact on the wind profiles mainly through the mean molecular mass $\mu(X_i)$.
  However, given that the total change is $\frac{\Delta \mu}{\mu} \sim 2\%$ for the duration of the wind (i.e. $ \lesssim 0.03\%$ for the typical $\Delta t_d$), we can safely assume that this is a valid approximation. 
  \footnote{
    Changes in opacity may also be relevant in principle which, at such high temperature, can be approximated with the formula given by \cite{Paczynski1983}, giving $\frac{\Delta \kappa}{\kappa}=\frac{\Delta X}{(1+X)}$. 
    In the models analyzed, this value is $\sim 0.4\%$ for the whole duration of the wind.
  }

\begin{figure}
  \centering 
  \includegraphics[keepaspectratio=true,width=0.95\columnwidth,clip=true,trim=0pt  0pt 0pt 40pt]{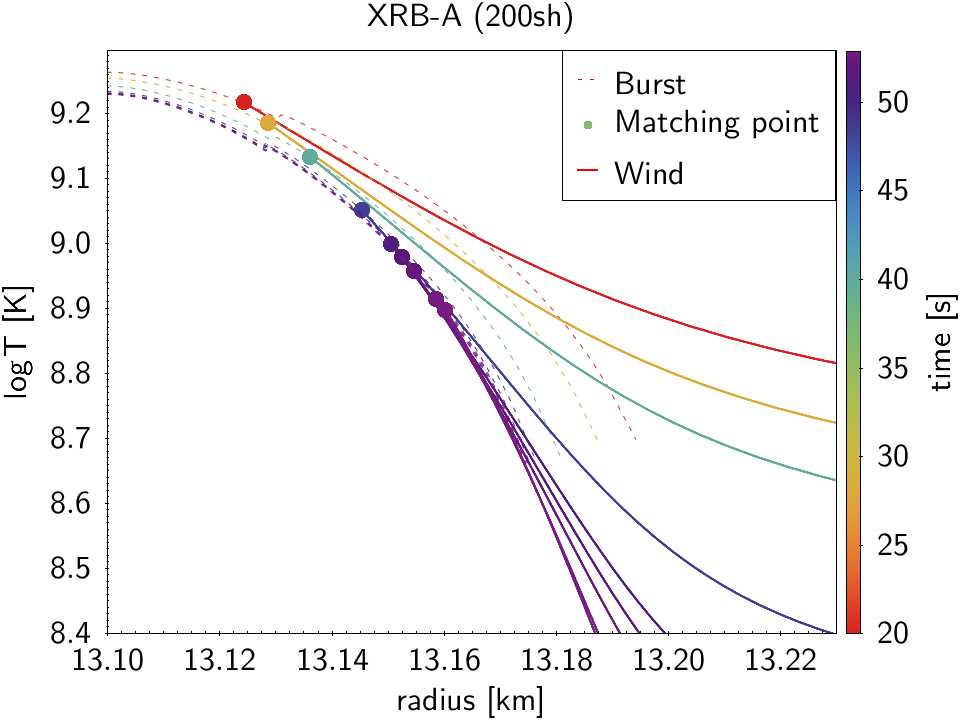} 
  \break 
  \includegraphics[keepaspectratio=true,width=0.95\columnwidth,clip=true,trim=0pt  0pt 0pt 40pt]{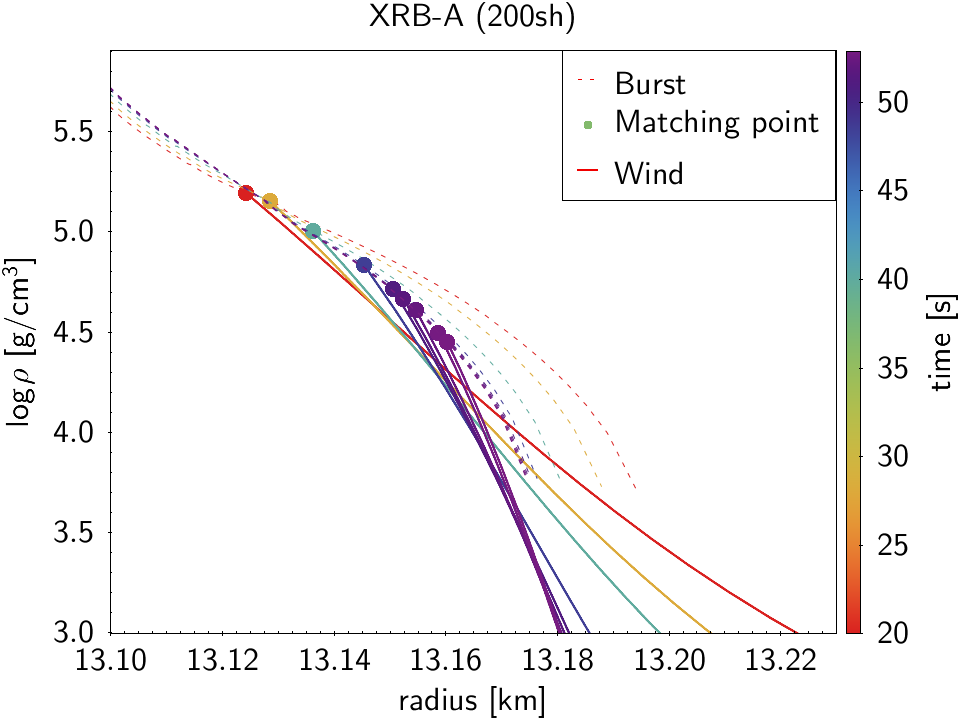}
  \caption{ 
    Detailed match of temperature (top) and density (bottom) radial profiles between wind (continuous lines) and hydrodynamic model (dashed lines) in  XRB-A.
    Burst-Wind profiles matching points are indicated with big circles. 
    Color scale indicates the time since burst peak expansion.
  }
  \label{fig: XRB-Wind T detail} 
\end{figure}

  A detailed view of the match between temperature profiles from wind and XRB hydrodynamic models is displayed in Fig. \ref{fig: XRB-Wind T detail}. 
  There, one can graphically appreciate not only the precise match in temperature at different times, but also the fact that the temperature gradient in both models has a similarly continuous value. 
  This feature is expected if convection is negligible, making radiative diffusion the only relevant term in the energy transport equation. 
  However, the same can not be said for the density match. Despite the match in density being more accurate in most cases, the density gradient is not continuous. 
  A density gradient appears in the mass conservation equation, and this discontinuity may originate in a significant contribution to integrated mass outflow from the time-dependent term. 
  Alternatively, it may derive from the different choice of surface boundary conditions in XRB hydrodynamic models, $P = P_\text{Rad}$, usual for non-expanded envelopes, and the fact they do not admit wind related mass-loss.
  These conditions can shape radial profiles differently, especially in upper shells close to the boundary, where the discontinuity in density gradient seems to be more prominent.

\subsection{Mass loss}
\label{sect: XRB-wind results mass-loss}

  After determination of the sequence of the matching solutions, between wind and hydrodynamic X-ray burst models, we can take a look at the chemical composition present at the matching points. 
  This will be, under our working hypothesis, the composition of the wind. 
  Every isotope present in it would ride this wind and escape the neutron star gravitational potential into the interstellar medium.
  
  In order to calculate the total mass-loss for each species, $\Delta m_i $, we need to integrate the mass outflow rate of every isotope, $\dot m_i(t)$, with respect to time, and over the duration of the wind.
  That is:
\begin{align}
  \label{eq: mass integral}
  \Delta m_i = \int_{t_o}^{t_f} \dot m_i(t) \, dt = \int_{t_o}^{t_f} \dot M(t) X_i(t) \, dt
\end{align}
  where $\dot M(t)$ is the total mass outflow as a function of time, 
  obtained as matching parameter of the wind profile at each time,
  and $t_o$ and $t_f$ are the initial and final times of the wind. 
  
\begin{figure}
  \centering 
  \includegraphics[keepaspectratio=true,width=0.95\columnwidth,clip=true,trim=0pt 0pt 0pt 40pt]{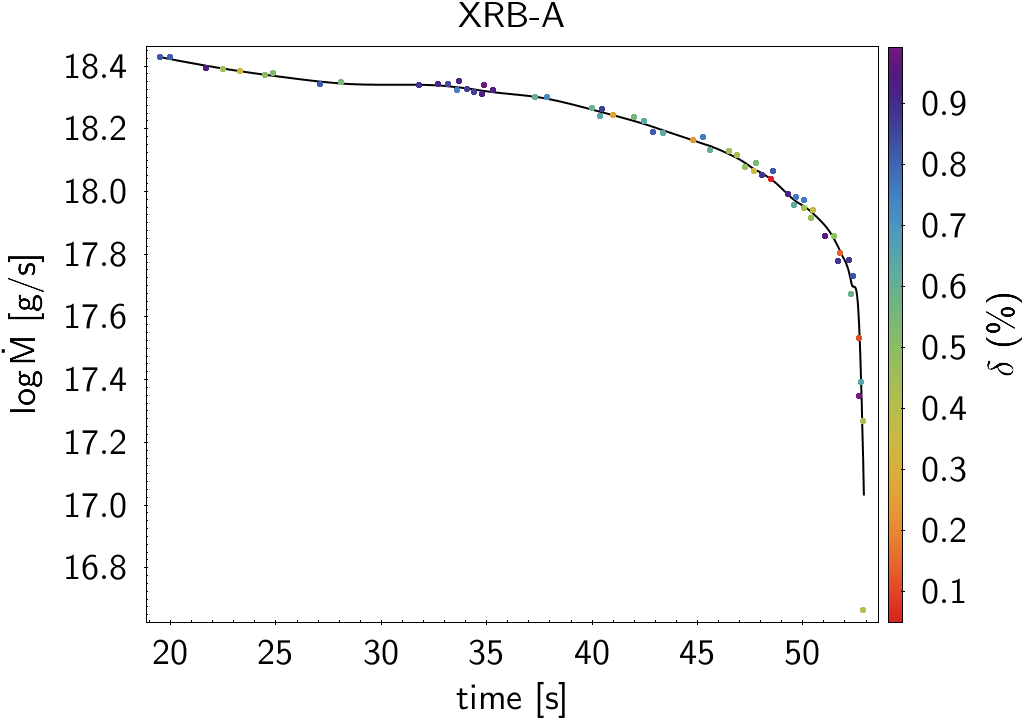} 
  \break 
  \includegraphics[keepaspectratio=true,width=0.95\columnwidth,clip=true,trim=0pt 0pt 0pt 40pt]{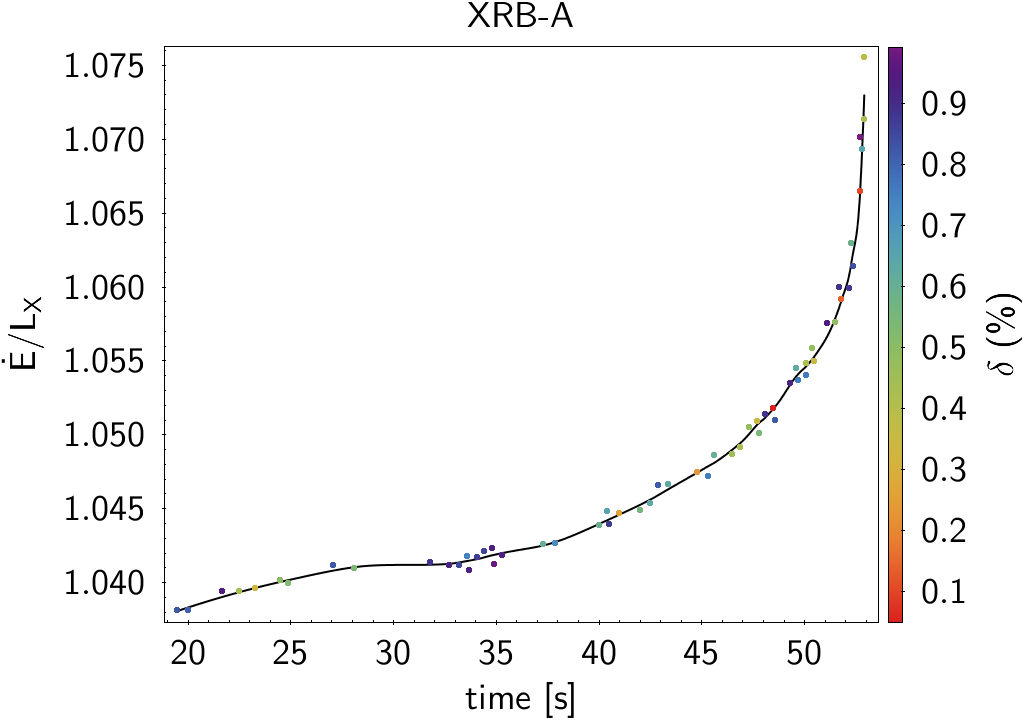} 
  \caption{ 
    Time evolution of mass outflow, $\dot M$ (top), and energy outflow, $\dot E$ 
    (bottom), in model XRB-A, and predictive curves using smoothing-interpolating technique (see text).
    Wind-burst matching data points are indicated with dots, and predicted values with a line. Color scale indicates the matching error 
    as defined by Eq. \ref{eq: W2B match error total}. 
  }
  \label{fig: XRB-Wind WiMP time smoothing} 
\end{figure}
  
  Although the accuracy of the match was relatively high in model XRB-A, with a significantly low matching error of $\delta < 0.01$, the matching points still show some degree of fluctuation in most physical variables and irregularity in time distribution (see Fig \ref{fig: XRB-Wind WiMP time smoothing}, for instance). 
  A smoothing technique using local regression was applied, incorporating the matching error as a weight factor in the form $\omega = \delta^{-2}$, so that the smoothing favors points with smaller error.
  This fixes the fluctuation in densely populated time intervals.
  Then, interpolation was used to reconstruct a curve. This serves two purposes: first, it fills up time gaps where sequences of potential matching points would lie in-between shells; and secondly, to get a large enough amount of equally distributed points to better approximate the integral in equation \ref{eq: mass integral}.
  Figure \ref{fig: XRB-Wind WiMP time smoothing} shows both the matching point data and this smooth-interpolating predictive curve, for the mass outflow, $\dot M$, and total energy outflow, $\dot E$.

\begin{figure}
  \centering 
  \includegraphics[keepaspectratio=true,width=0.95\columnwidth,clip=true,trim=0pt 0pt 0pt 40pt]{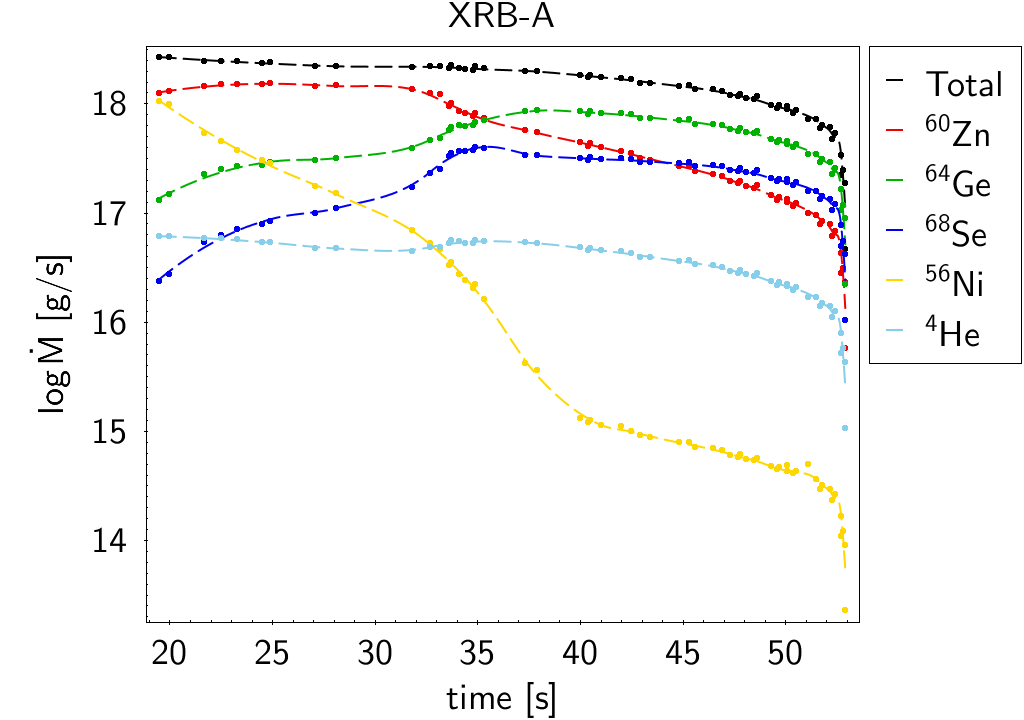}
  \break 
  \includegraphics[keepaspectratio=true,width=0.95\columnwidth,clip=true,trim=0pt 0pt 0pt 40pt]{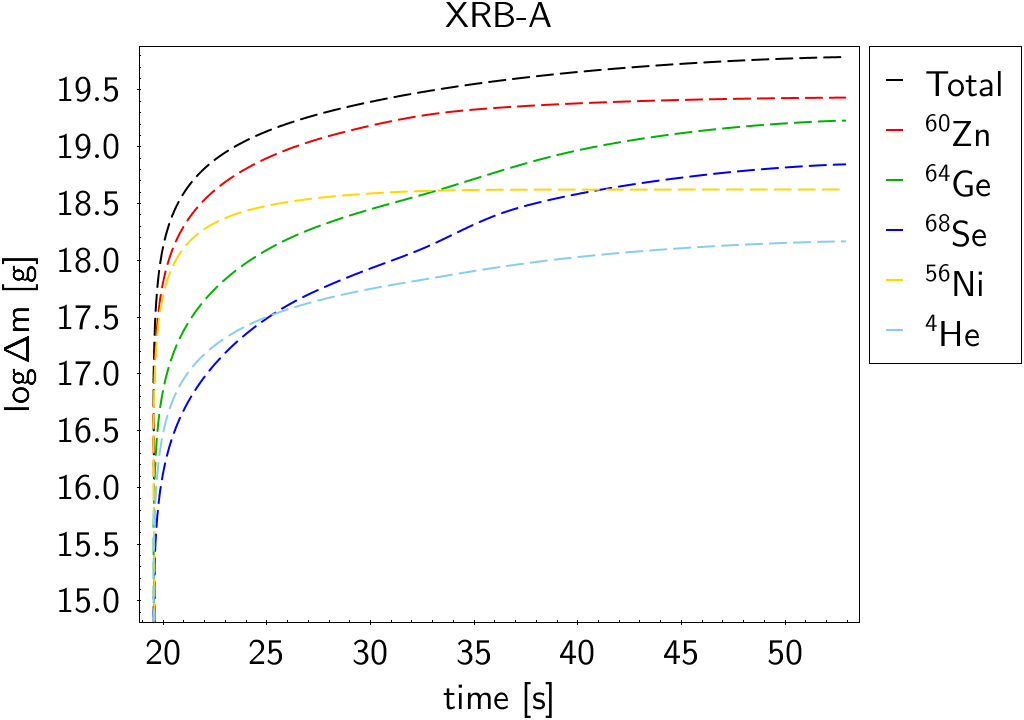}
  \caption{ 
    Top: Time evolution of mass outflow ($\dot M$) for the top 5 species (by ejected mass) and total mass outflow. 
    Wind-Burst model matching points are indicated with dots, the predictive curve (dashed line) is also shown for each of them. 
    Bottom: Time-integrated ejected mass ($\Delta m$) for the same isotopes. 
    Both panels correspond to model XRB-A.
  }
  \label{fig: XRB-Wind mass time-integrated} 
\end{figure}

  Figure \ref{fig: XRB-Wind mass time-integrated} shows both the predicted time-evolution curves of mass outflow rates for the top five species, and their time-integrated ejected mass. 
  The total mass outflow seems to increase towards earlier times than the ones shown. The matching technique could still be applied, in principle, to find earlier matching points, but this was dismissed due to hypothesis inconsistency between models (see Sect. \ref{sect: XRB-wind match analysis}).
  However, the total mass outflow is expected to go back down by the time of the wind actual onset. 
  The total mass yield obtained is $\Delta m = 6.16\E{19}\ \texttt{g} $,
  for the time interval analyzed.\footnote{
    That is, without considering any contribution from possible earlier matching points. 
    In that case, a quick extrapolation for the estimated mass yield could be twice as much.
  }
  Considering that model XRB-A has a recurrence time of $5.9\texttt{ hr}$, and assuming no period of inactivity (i.e., an upper limit to its real contribution), this implies a yearly output of $4.6 \E{-11}\,\Msun /\texttt{yr}$, that is $2.6\%$ of the mass-accretion rate.

\begin{figure}
  \centering 
  \includegraphics[keepaspectratio=true,width=\columnwidth,clip=true,trim=0pt  0pt 0pt 35pt]{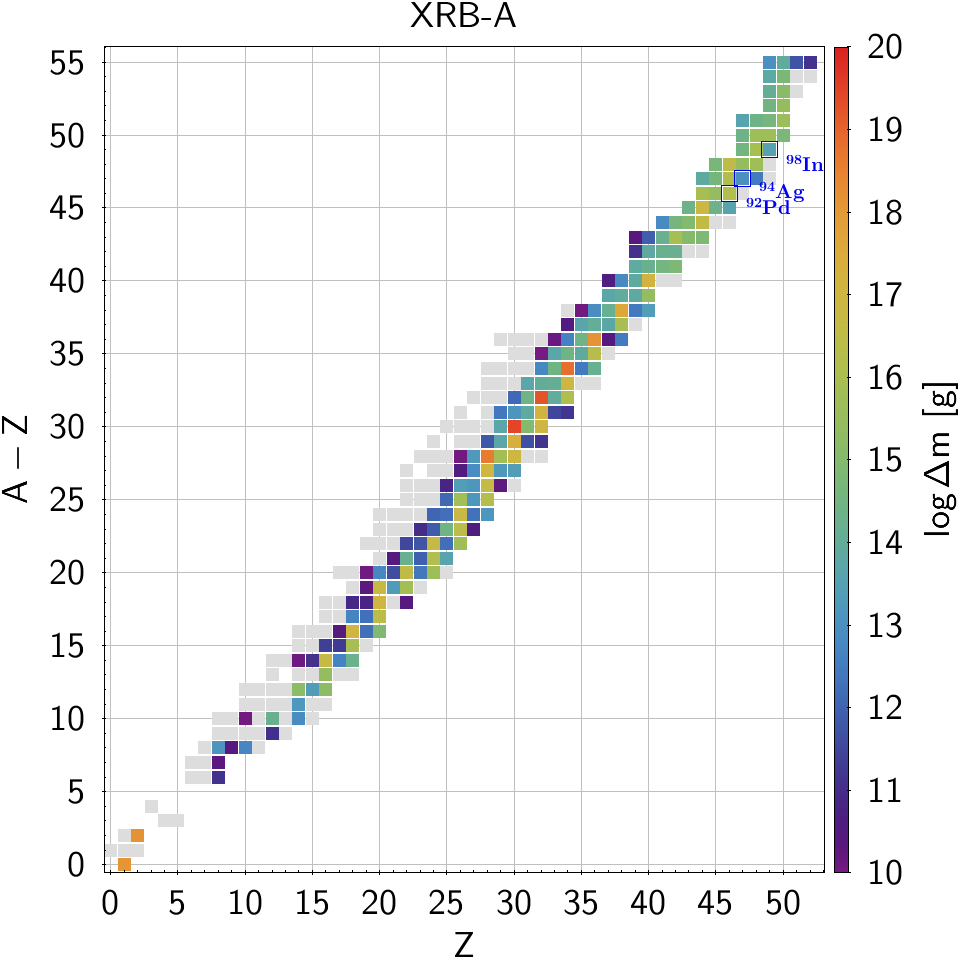} \hfill
  \caption{
    Mass ejected per isotope for the overall duration of the wind, from model XRB-A. 
    Color scale indicates the mass ejected only for isotopes for which $\log \Delta m > 10$, the rest is gray. Some species of interest are also marked.
  }
  \label{fig: XRB-Wind isotopes A-Z} 
\end{figure}

\begin{figure}
  \centering 
  \includegraphics[keepaspectratio=true,width=\columnwidth,clip=true,trim=0pt  0pt 0pt 35pt]{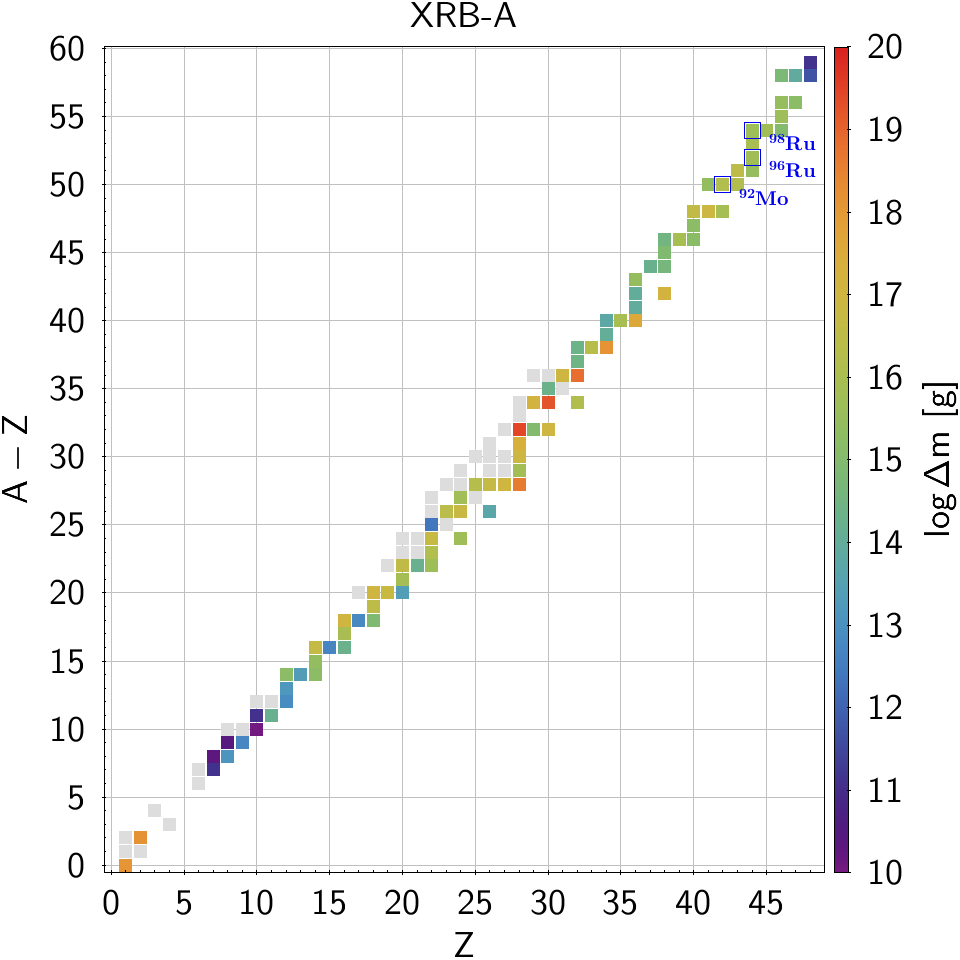}
  \caption{
    Stable isotopes mass yield from stellar wind in model XRB-A. 
    This corresponds to the final products from radioactive decay of the unstable isotopes shown in Fig \ref{fig: XRB-Wind isotopes A-Z}.
  }
  \label{fig: XRB-Wind isotopes A-Z stable} 
\end{figure}

  Figure \ref{fig: XRB-Wind isotopes A-Z} shows the total mass yield of every isotope produced in the XRB and ejected by the stellar wind.
  Mass yield is color-coded only for species with a mass output $\Delta m_i > 10^{10} \texttt{ g}$. 
  Most of the isotopes produced in the XRB are unstable and will eventually decay (mainly through $\beta$ process) into stable ones. 
  The final resulting stable isotopes were also computed
  and are shown in Fig \ref{fig: XRB-Wind isotopes A-Z stable}.
  Values for the top ten mass-yielding isotopes are shown in Table \ref{tab: top isotopes}. 
  Almost $90\%$ of the total expelled mass is constituted by the first four isotopes listed in the table:
  $^{60}\text{Ni}$, $^{64}\text{Zn}$, $^{68}\text{Ge}$, $^{56}\text{Ni}$.
  All ten species shown in the table account for over $97\%$ of the overall mass ejected.

\begin{table}
  \begin{center}
  \caption{
    Top ten stable isotopes (after radioactive decay) by mass yield, from stellar wind in model XRB-A.
  }
  \label{tab: top isotopes}
  \small
%
  \begin{tabular}[t]{|lrrrr|}
    \hline
	Sym     &   Z  &   A  &   Mass (g)                  &    X            \\
    \hline
    Ni      &  28  &  60  &   $ 2.71 \E{ 19} $   &   $ 4.39 \E{-1} $    \\
    Zn      &  30  &  64  &   $ 1.70 \E{ 19} $   &   $ 2.76 \E{-1} $    \\
    Ge      &  32  &  68  &   $ 6.99 \E{ 18} $   &   $ 1.13 \E{-1} $    \\
    Ni      &  28  &  56  &   $ 4.20 \E{ 18} $   &   $ 6.82 \E{-2} $    \\
    He      &   2  &   4  &   $ 1.47 \E{ 18} $   &   $ 2.38 \E{-2} $    \\
    Se      &  34  &  72  &   $ 1.42 \E{ 18} $   &   $ 2.31 \E{-2} $    \\
    H       &   1  &   1  &   $ 1.29 \E{ 18} $   &   $ 2.09 \E{-2} $    \\
    Kr      &  36  &  76  &   $ 4.04 \E{ 17} $   &   $ 6.56 \E{-3} $    \\
    Ni      &  28  &  59  &   $ 2.11 \E{ 17} $   &   $ 3.43 \E{-3} $    \\
    Cu      &  29  &  63  &   $ 1.48 \E{ 17} $   &   $ 2.41 \E{-3} $    \\
    \hline
  \end{tabular}
  \end{center}
  \normalsize
\end{table}

  It has been suggested that if a tiny fraction of the accreted envelope is ejected through radiation-driven winds, XRBs may potentially be the source of some light p-nuclei, such as $^{92,94}\text{Mo}$ and $^{96,98}\text{Ru}$ \cite[see][]{Schatz_1998, Schatz_2001}, which are systematically under-produced in all canonical scenarios proposed to date for the synthesis of such species.
  With the results found here we can make a rough estimate regarding the possible contribution of p-nuclei. 
  Their progenitor ($^{92}\text{Pd}$, $^{94}\text{Ag}$ and $^{98}\text{In}$) and final stable species available are all marked in Figs. \ref{fig: XRB-Wind isotopes A-Z} and \ref{fig: XRB-Wind isotopes A-Z stable}, respectively.
  The significant  mass yields were: 
  $\Delta m_{92\text{Mo}} = 1.57\E{16} \texttt{ g}$, 
  $\Delta m_{96\text{Ru}} = 5.75\E{15} \texttt{ g}$,
  and
  $\Delta m_{98\text{Ru}} = 4.72\E{15} \texttt{ g}$,
  for the duration of the wind, which corresponds to mass fractions of:
  $X_{92\text{Mo}} = 2.57\E{-4}$,
  $X_{96\text{Ru}} = 9.33\E{-5}$,
  and
  $X_{98\text{Ru}} = 7.67\E{-5}$. 
  With a burst recurrence time of $\tau_{rec} = 5.9\texttt{ hr}$, this gives annual yields of:
  $ 1.18 \E{-14} \,\Msun/\texttt{yr}$,
  $ 4.29 \E{-15} \,\Msun/\texttt{yr}$, and  
  $ 3.53 \E{-15} \,\Msun/\texttt{yr}$, respectively.

\begin{table*}[htp]
  \centering
  \caption{
    Estimates of light p-nuclei isotopes contribution from XRB-A wind to the galactic abundances (see Sect. \ref{sect: XRB-wind results mass-loss}). 
  }
  \label{tab: galactic isotopes}
  \begin{tabular}{|l|lll|}
    \hline
    Isotope                                        &   $^{92}$Mo               &    $^{96}$Ru                 &  $^{98}$Ru      \\
    \hline
    XRB-A wind mass yield per burst  (g)           &   $ 1.57 \E{ 16} $  &    $ 5.75 \E{ 15} $    &  $ 4.73 \E{ 15} $   \\
    XRB-A annual yield (M$_\odot$/yr)              &   $ 1.18 \E{-14} $  &    $ 4.29 \E{-15} $    &  $ 3.53 \E{-15} $   \\
    XRB-A wind mass fraction $X_i$                 &   $ 2.57 \E{-04} $  &    $ 9.33 \E{-05} $    &  $ 7.67 \E{-05} $   \\
    XRB-A life-time contribution (M$_\odot$)
			   \tablefootmark{a}       &   $ 5.88 \E{-05} $  &    $ 2.15 \E{-05} $    &  $ 1.76 \E{-05} $   \\
    \hline
    Mass fraction $X_{i}$ in solar system                           &   $ 9.27 \E{-10} $  &    $ 2.59 \E{-10} $    &  $ 8.80 \E{-11} $   \\
    Total isotope mass in Milky Way (M$_\odot$)
			  \tablefootmark{b}        &   $ 1.85 \E{ 02} $  &    $ 5.17 \E{ 01} $    &  $ 1.76 \E{ 01} $   \\
    \hline
    Sources like XRB-A needed to match             &   $ 3.15 \E{ 06} $  &    $ 2.41 \E{ 06} $    &  $ 9.98 \E{ 05} $   \\
\hline
  \end{tabular}
\tablefoot{
  \tablefoottext{a}{
    XRB-A is considered to be active for a period of $5 \texttt{ Gyr}$ (see text).
  }
  \tablefoottext{b}{
    Total galactic mass used for calculations is $2\E{11} \,\Msun$ (see text).
  }
}
\end{table*}

  The remaining question is how significant is the XRB wind contribution we found for the observed Galactic abundance of these light p-nuclei. 
  Current measurements in primitive meteorites \cite[see][and references therein]{Lodders2009} report the following abundances: 
  $X^\odot_{92\text{Mo}} = 9.27\E{-10}$,
  $X^\odot_{96\text{Ru}} = 2.59\E{-10}$,
  and
  $X^\odot_{98\text{Ru}} = 8.80\E{-11}$.
  For an estimation of the XRB wind contribution to the Galactic abundances, the total net mass of each element ejected by the XRB winds must be compared to the total mass of that element present in the Galaxy.
  A raw estimate can be obtained in the following way (results are summarized in Table \ref{tab: galactic isotopes}). 
  The total observable mass of the Milky Way is believed to lie in the range of $[100-400] \E{9} \,\Msun$.\footnote{
    Here, we consider only baryonic matter, no dark matter estimate is included \cite[see, e.g.,][]{Kafle++2014,McMillan2017,Watkins++2019}.
  }
  Assuming a value of $200\E{9}\ \Msun$, and the mass fractions inferred from meteoritic samples in our solar system, we get the following estimates for the amount of those p-nuclei present in our Galaxy:
  $\Delta m^{\texttt{MW}}_{92\text{Mo}} = 185 \ \Msun$, 
  $\Delta m^{\texttt{MW}}_{96\text{Ru}} = 51.7 \ \Msun$,
  and
  $\Delta m^{\texttt{MW}}_{98\text{Ru}} = 17.6 \ \Msun$, where the ``MW'' superscript stands for Milky Way.
  Now, assuming that all XRB sources contribute with a similar chemical abundance pattern than model XRB-A, and assuming no periods of inactivity (i.e., all sources experience one burst after 5.9 hr, continuously 
  throughout their entire lifetime),
  we can get an upper limit of the overall mass contribution for each species to the galactic abundance, by simply multiplying the annual yields by the estimated lifetime of X-ray binaries ($5\texttt{ Gyr}$).\footnote{
    This is a traditionally assumed value for X-ray binaries lifetime, but it is currently not very well constrained. 
    Other works suggest much shorter lifetimes, by even 3 orders of magnitude \cite[see][]{Iben++1995,Podsiadlowski++2002,Pfahl++2003}. 
    In such case, our final conclusion (see further ahead) is strengthened, since even more XRB sources would be needed to explain the observed mass abundances. 
  }
  This results in:
  $\Delta m^{\texttt{XRB}}_{92\text{Mo}} = 1.59\E{-4} \ \Msun$, 
  $\Delta m^{\texttt{XRB}}_{96\text{Ru}} = 5.8\E{-5} \ \Msun$,
  and
  $\Delta m^{\texttt{XRB}}_{98\text{Ru}} = 4.76\E{-5} \ \Msun$.
  The ratio between the mass of each species present in the Galaxy and the amount ejected by our typical XRB source throughout its life-time (i.e. $\Delta m^\texttt{MW}_i / \Delta m^\texttt{XRB}_i $) yields an estimate of the number of XRB sources similar to XRB-A needed to account for the origin of these p-nuclei. 
  This ratio gives different results for each isotope. 
  Between a few hundred thousand to over a million sources like XRB-A are needed to match. 
  In contrast, the total number of identified XRB sources so far is around one hundred.\footnote{
    See \url{http://www.sron.nl/~jeanz/bursterlist.html} for an updated list of known galactic type I XRB systems.
  }
  That is, assuming that XRBs are the only source for these species and that our XRB-A is a representative case, then $10^3-10^4$ times more sources than the ones already discovered would be needed to explain the observed mass abundances of these light p-nuclei.
  This suggests that XRBs are unlikely to be the main origin of these isotopes.

\subsection{Observables, correlations and other physical magnitudes}
\label{sect: XRB-wind results observables}

  Another important aspect we would like to analyze from the XRB-wind matching results is the evolution of observable features and how they relate to other predicted physical variables.

\begin{figure*}
  \centering 
  \includegraphics[keepaspectratio=true,width=0.49\textwidth,clip=true,trim=0pt  0pt 0pt 40pt]{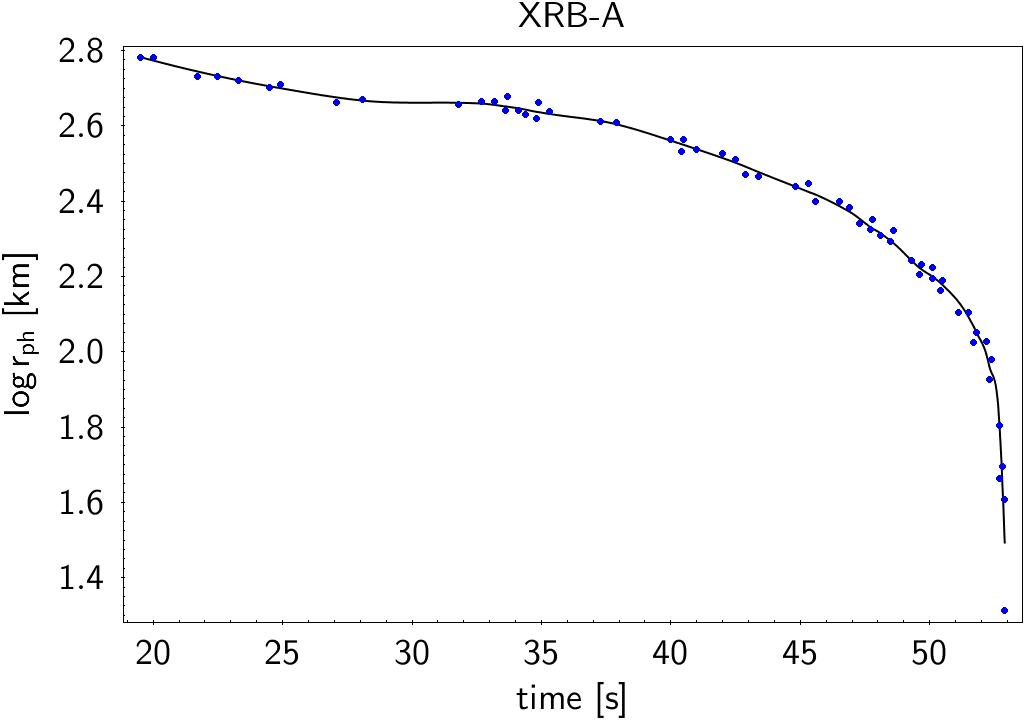}
  \hfill
  \includegraphics[keepaspectratio=true,width=0.49\textwidth,clip=true,trim=0pt  0pt 0pt 40pt]{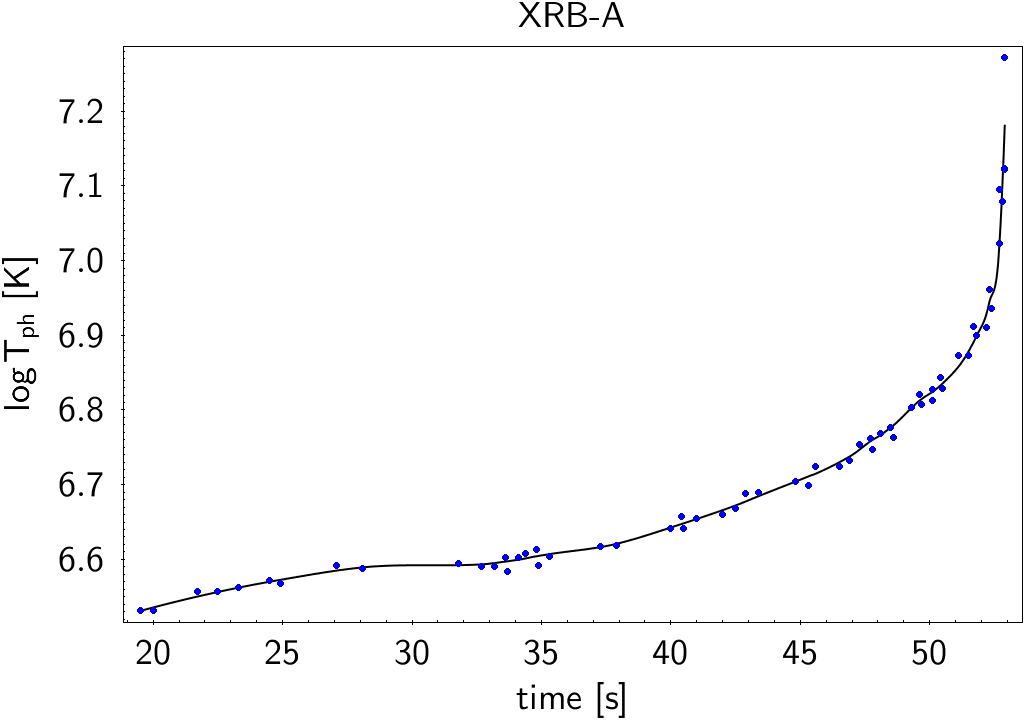} 
  \linebreak \linebreak
  \includegraphics[keepaspectratio=true,width=0.49\textwidth,clip=true,trim=0pt  0pt 0pt 40pt]{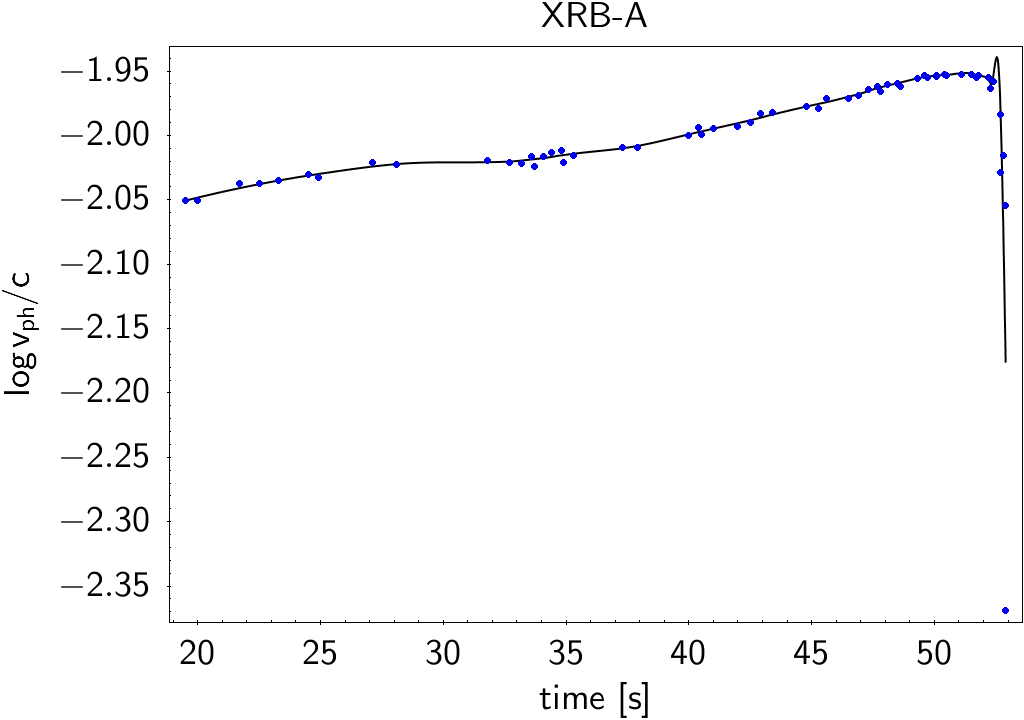} 
  \hfill
  \includegraphics[keepaspectratio=true,width=0.49\textwidth,clip=true,trim=0pt  0pt 0pt 40pt]{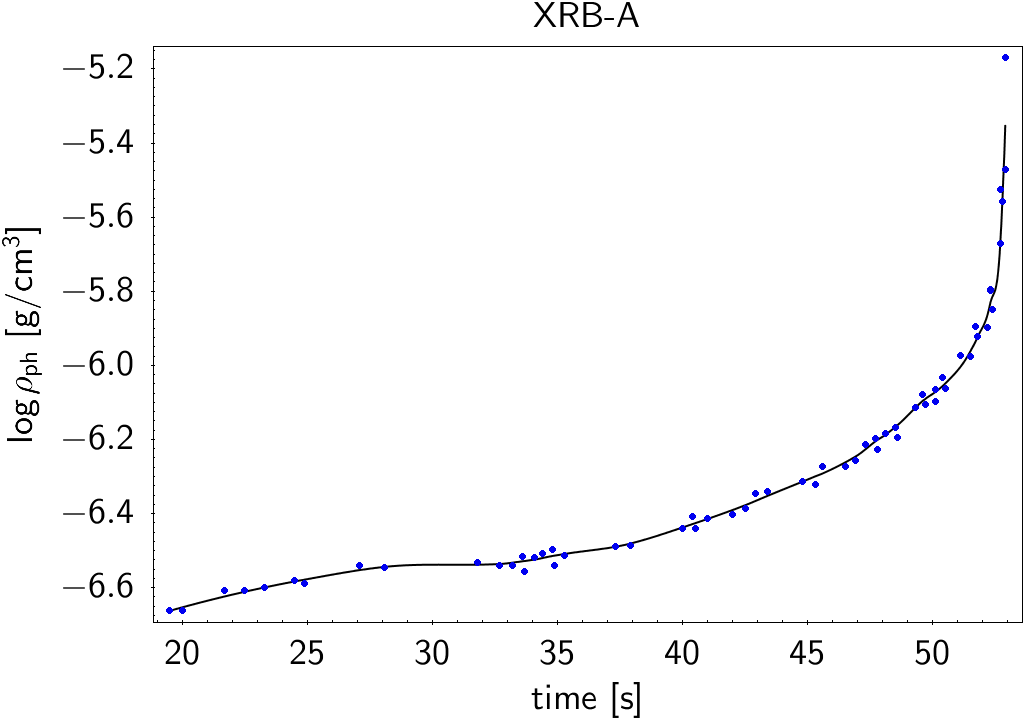}  
  \linebreak \linebreak
  \includegraphics[keepaspectratio=true,width=0.49\textwidth,clip=true,trim=0pt  0pt 0pt 40pt]{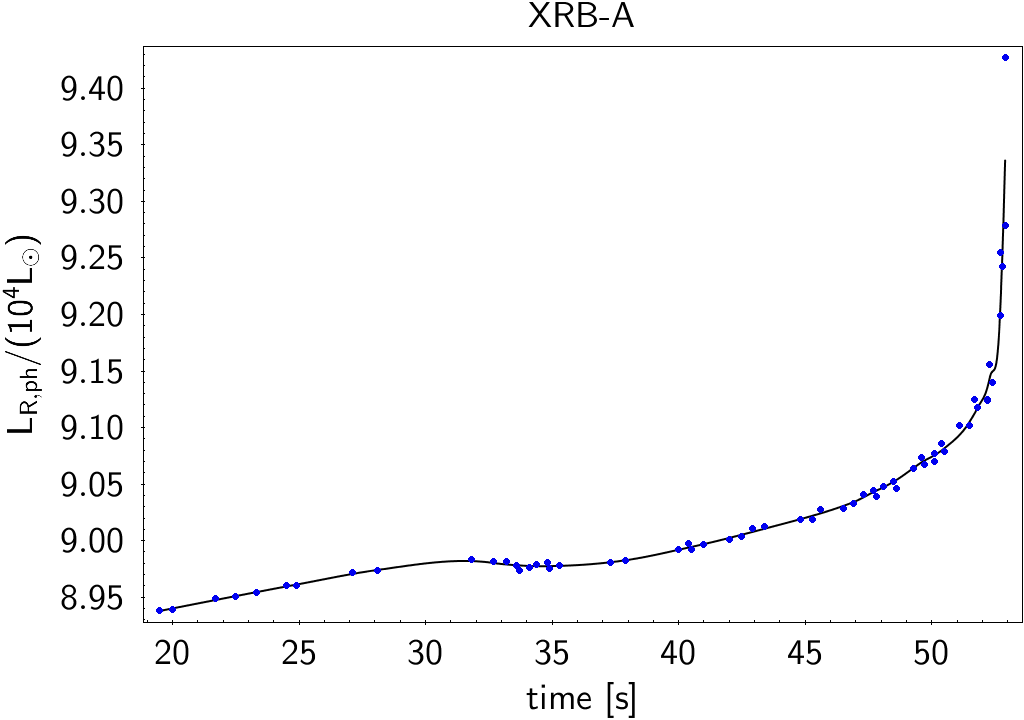}  
  \hfill
  \includegraphics[keepaspectratio=true,width=0.49\textwidth,clip=true,trim=0pt  0pt 0pt 40pt]{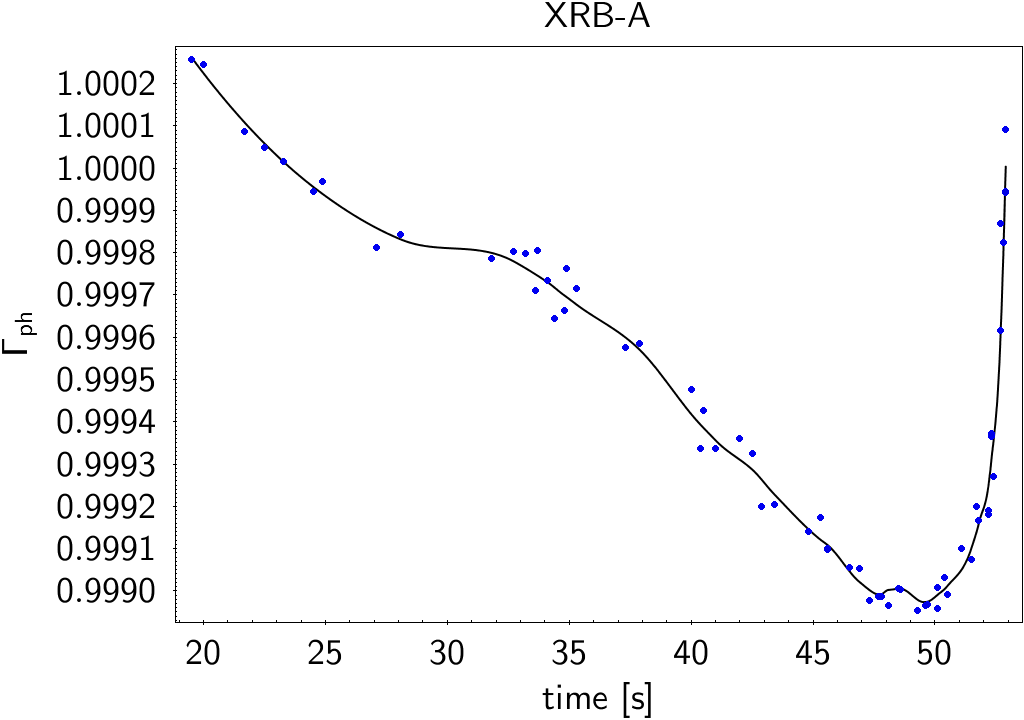}  
  \caption{ 
    Time evolution of photospheric magnitudes, in model XRB-A. 
    In reading order, radius, temperature, wind velocity, density, radiative luminosity and its ratio to local Eddington luminosity. 
    Values corresponding to matching wind profiles are indicated with dots, and predicted values (using smoothing-interpolating technique) with a line. 
    }
    \label{fig: XRB-Wind photosphere time smoothing} 
\end{figure*}

  Time-evolution of photospheric magnitudes in XRB-A is shown in Fig. \ref{fig: XRB-Wind photosphere time smoothing}. 
  Photospheric radial expansion is receding from over $600 \texttt{ km}$, steadily for most of the time interval displayed, and drops rapidly in the last few seconds, seemingly towards the final XRB-wind matching radius ($\sim 13.17\texttt{ km}$), close to the NS core surface. 
  Effective temperature appears to evolve in an exactly inverse manner, rising from around $4\E{6}\texttt{ K}$ in the earlier moments where the wind atmosphere is most expanded, and ending in a quick spike ($> 1.6 \E{7} \texttt{ K}$) that seems to sharply rise up towards the much higher photospheric temperature ($\sim 5\E{8} \texttt{ K}$) of the envelope after the wind phase.
  Wind velocity at the photosphere varies little around $\sim 0.01 \, c$ for the most part, and drops abruptly by the end.
  Density shows an almost identical behavior to temperature. 
  
  Photospheric luminosity rises slowly through most of the wind and spikes towards the end, in a similar way to temperature.
  Analysis of the full wind evolution from lower resolution models \cite[i.e., Model 1 in][]{Jose2010}
  hint that this rise may in fact be the recovery part of an overall small dip in luminosity ($\sim 5-6\%$) with respect to the average luminosity while the wind is active ($ L_X \sim 3.52\E{38} \  \texttt{erg/s} $).
  This average luminosity is in turn much smaller than the values typically reported in XRB hydrodynamic simulations, especially if the wind phase appears during their peak expansion.
  The onset of the stellar wind is expanding the envelope further, cooling it down towards the photosphere. 
  This in turn rises opacity and allows for the absorption of more radiative energy into the gas that is transformed into kinetic energy, making the photospheric luminosity drop. 
  This drop in photospheric luminosity during the wind phase with respect to the wind-less envelope could, in principle, be observed as a flattening of the luminosity peak for the duration of the wind or, if the drop is significant and the wind short-lived, as a double peak.
  From the observational point of view, such a feature, could be a clear signal of the presence of a stellar wind.
  
  We noted the similarity between photospheric temperature and density time evolution curves, and their apparent inverse relation to photospheric radius. 
  In the characterization study shown in \cite{HSJ2020} we found similar correlations due to the fact that photospheric luminosity was close to Eddington luminosity. 
  That is:
\begin{align}
  T_\text{ph} \sim {r_\text{ph}}^{-1/2},
  \label{eq: correlation T r}
  \\
  \rho_\text{ph} \sim {r_\text{ph}}^{-1}.
  \label{eq: correlation rho r}
\end{align}
  We also stated the correlation between observable features, such as photospheric luminosity and wind velocity, and wind parameters determined by physical conditions of the inner parts of the envelope,
\begin{align}
    \dfrac{8}{3}\ \dfrac{v_\text{ph}}{c} =  \dfrac{GM}{r_\text{ph}} \dfrac{\dot M}{L_\text{R,ph}} \simeq \dfrac{\dot E }{L_\text{R,ph}}-1  .
    \label{eq: photospheric correlations with parameters}
\end{align}
  Now we are able to confirm that these correlations hold for the more realistic scenario studied here. 
  The regression results are shown in Table \ref{tab: Regression results XRB-A}.

\begin{table}[htp]
  \begin{center}
  \caption{Regression results from correlations among observable variables and wind parameters in model XRB-A.}
  \label{tab: Regression results XRB-A}
  \begin{tabular}{r|ccc}
    Regression model   &   a   &   c   & Correlation \\
    \hline
    $ \log T_\text{ph}    = \text{c} + \text{a} \log r_\text{ph} $    &   $-0.503$   &   $10.45$   & $0.9999$ \\
    $ \log \rho_\text{ph} = \text{c} + \text{a} \log r_\text{ph} $    &   $-1.013$    &   $1.22$   & $0.9999$ \\
    $ \frac{8}{3} \frac{v_\text{ph}}{c} = \text{c} + \text{a} \frac{\dot M}{L_{R,\text{ph}}}\frac{G M}{ r_\text{ph}} $    &   $0.997$    &   $5.5\E{-5}$   & $0.9999$ \\
    $  \frac{\dot E}{L_{R,\text{ph}}} -1 = \text{c} + \text{a} \frac{\dot M}{L_{R,\text{ph}}}\frac{G M}{ r_\text{ph}} $    &   $0.992$    &   $3.6\E{-4}$   & $0.9994$ 
  \end{tabular}
  \end{center}
\end{table}

\begin{figure*}
  \centering 
  \includegraphics[keepaspectratio=true,width=0.49\textwidth,clip=true,trim=0pt  0pt 0pt 40pt]{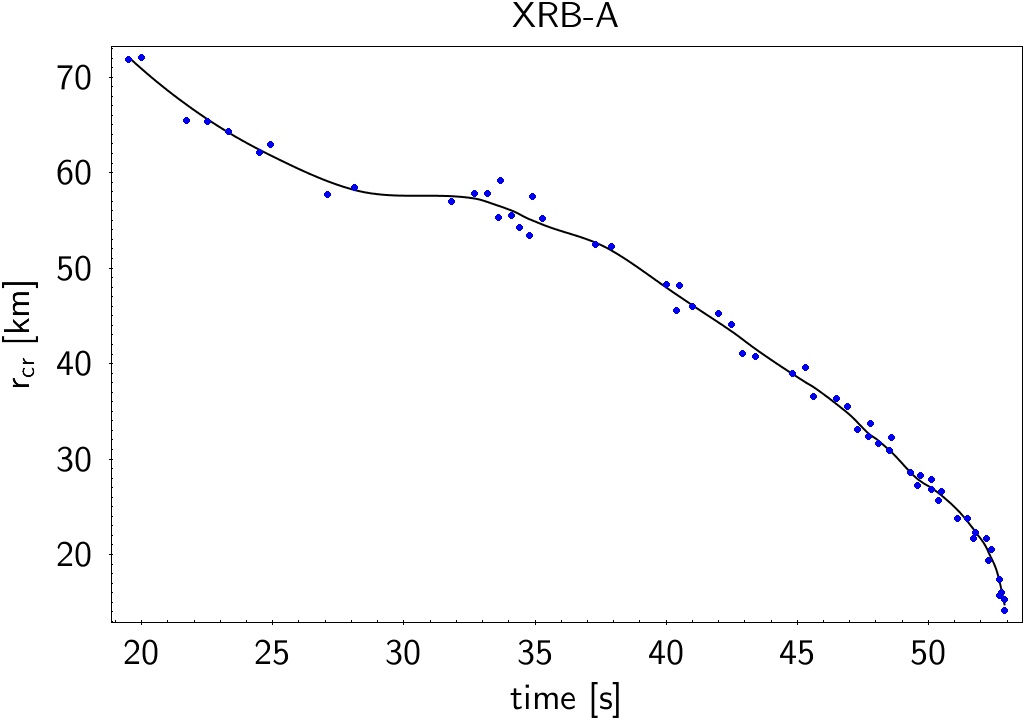}  
  \hfill
  \includegraphics[keepaspectratio=true,width=0.49\textwidth,clip=true,trim=0pt  0pt 0pt 40pt]{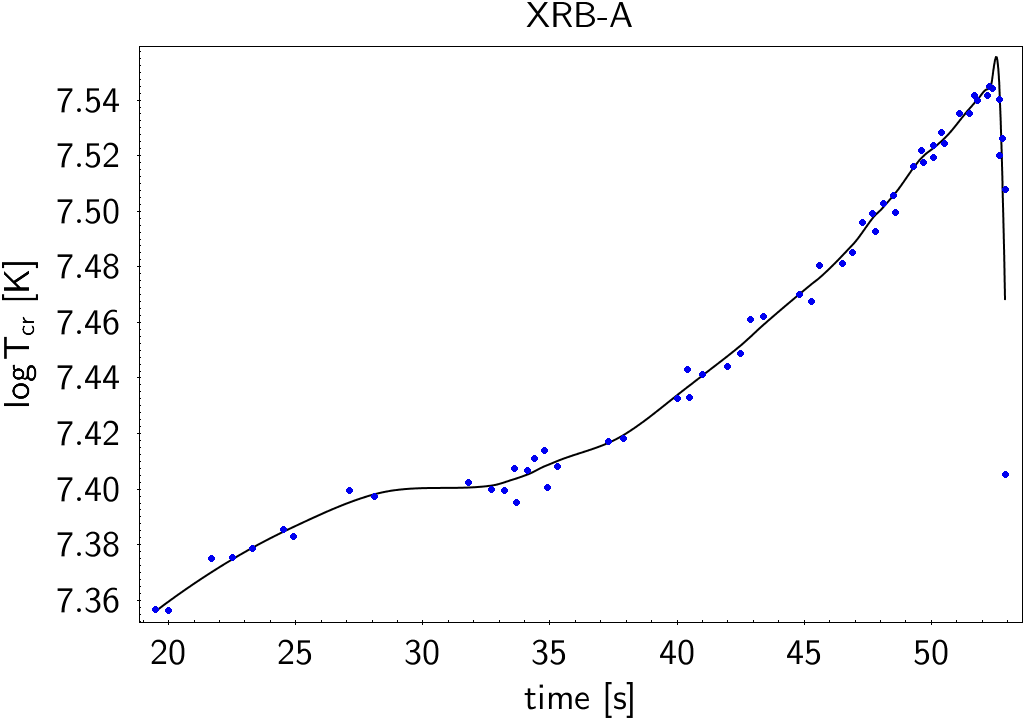} 
  \linebreak \linebreak
  \includegraphics[keepaspectratio=true,width=0.49\textwidth,clip=true,trim=0pt  0pt 0pt 40pt]{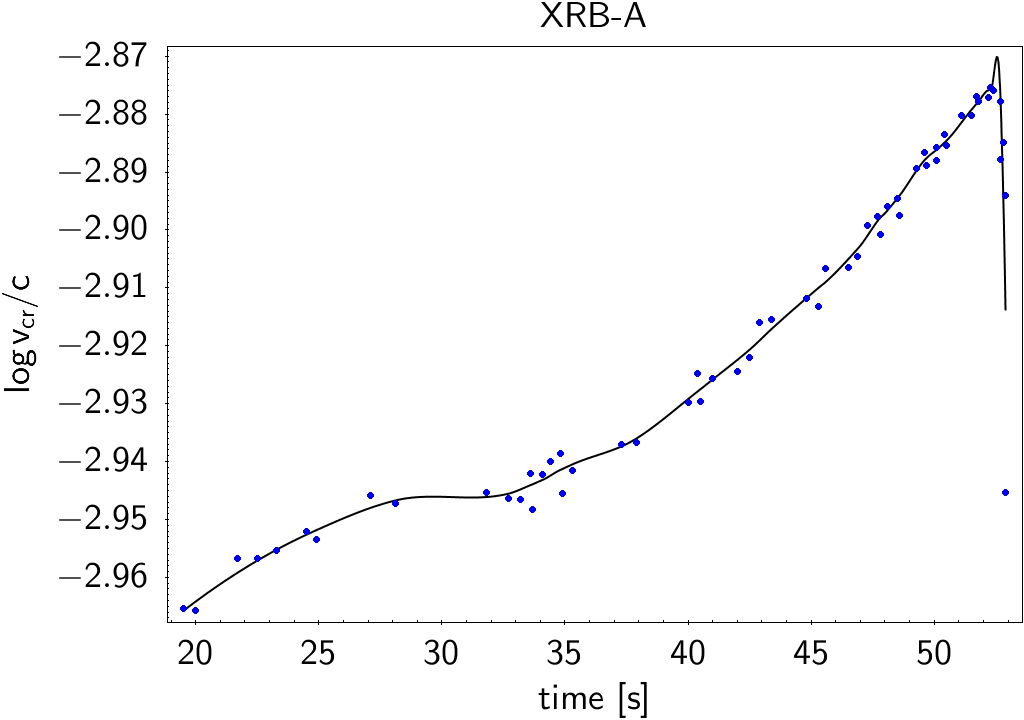} 
  \hfill
  \includegraphics[keepaspectratio=true,width=0.49\textwidth,clip=true,trim=0pt  0pt 0pt 40pt]{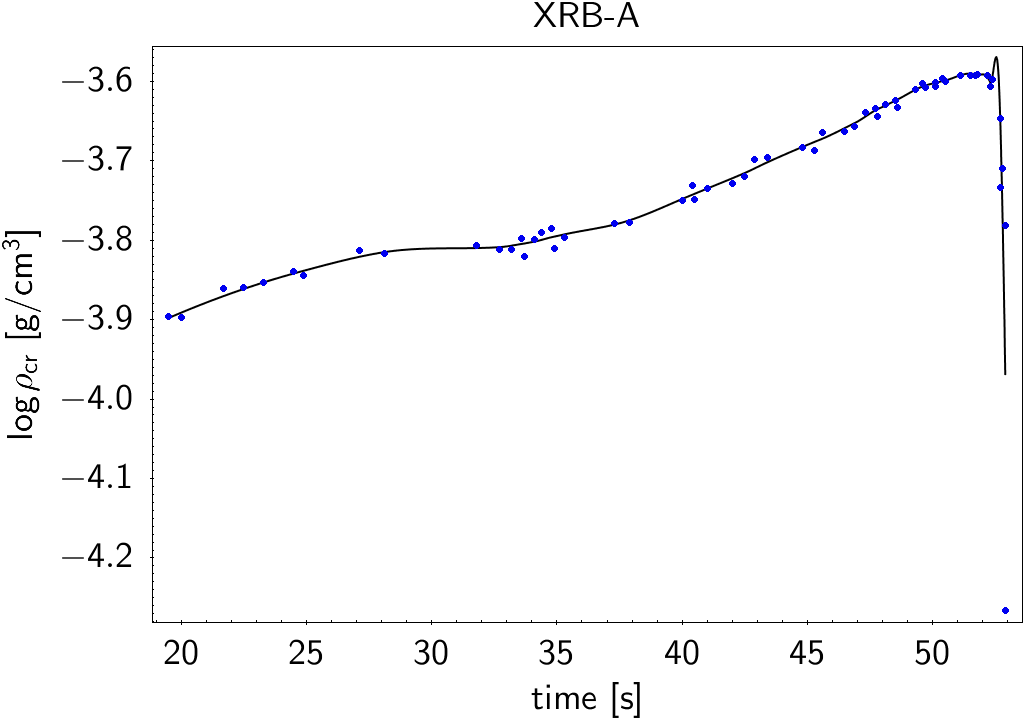} 
  \linebreak \linebreak
  \includegraphics[keepaspectratio=true,width=0.49\textwidth,clip=true,trim=0pt  0pt 0pt 40pt]{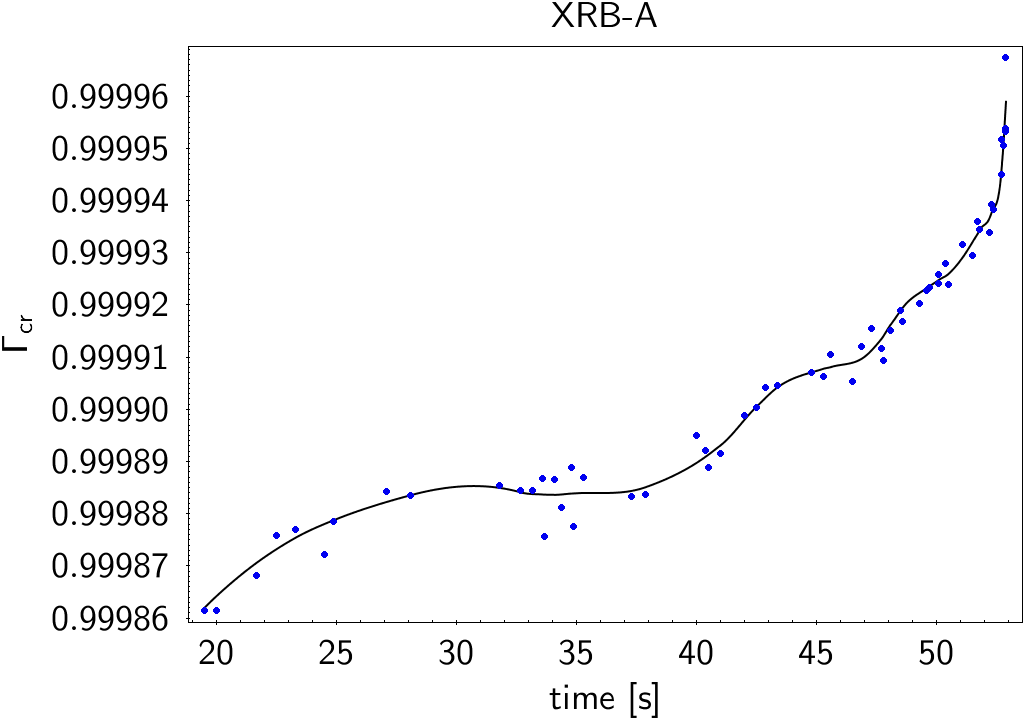}  
  \hfill
  \includegraphics[keepaspectratio=true,width=0.49\textwidth,clip=true,trim=0pt  0pt 0pt 40pt]{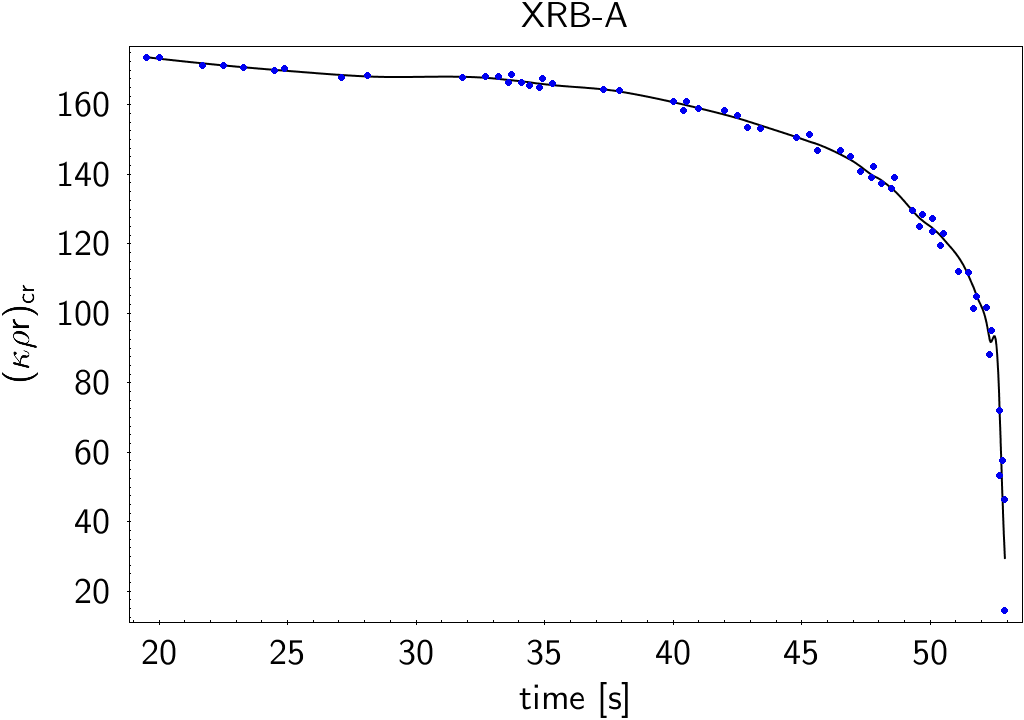}  
  \caption{ 
    Same as Fig. \ref{fig: XRB-Wind photosphere time smoothing}, for magnitudes evaluated at the critical point, in reading order: radius, temperature, density, 
    ratio of radiative luminosity to local Eddington limit, and effective optical depth. 
  }
  \label{fig: XRB-Wind critical time smoothing} 
\end{figure*}

\begin{figure*}
  \centering 
  \includegraphics[keepaspectratio=true,width=0.49\textwidth,clip=true,trim=0pt  0pt 0pt 40pt]{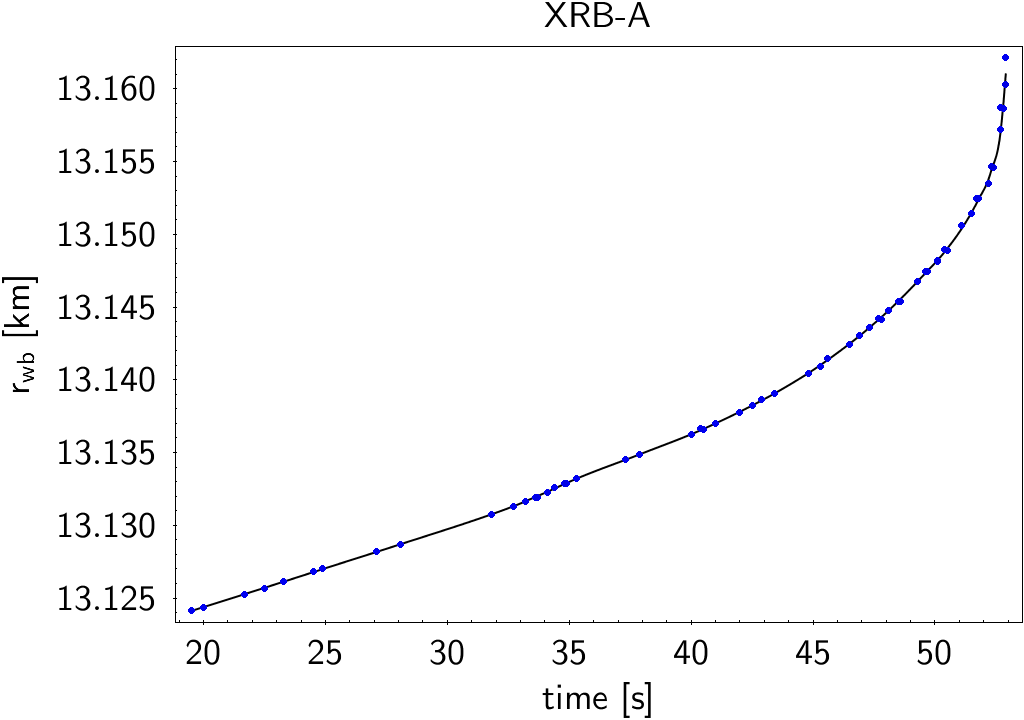}  
  \hfill
  \includegraphics[keepaspectratio=true,width=0.49\textwidth,clip=true,trim=0pt  0pt 0pt 40pt]{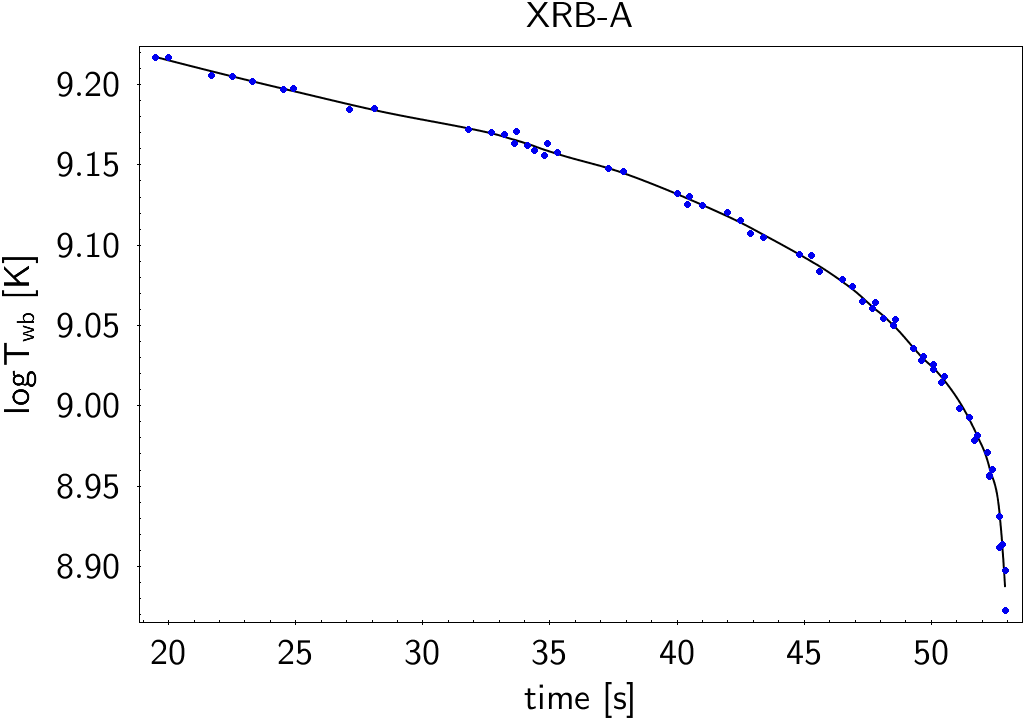} 
  \linebreak \linebreak
  \includegraphics[keepaspectratio=true,width=0.49\textwidth,clip=true,trim=0pt  0pt 0pt 40pt]{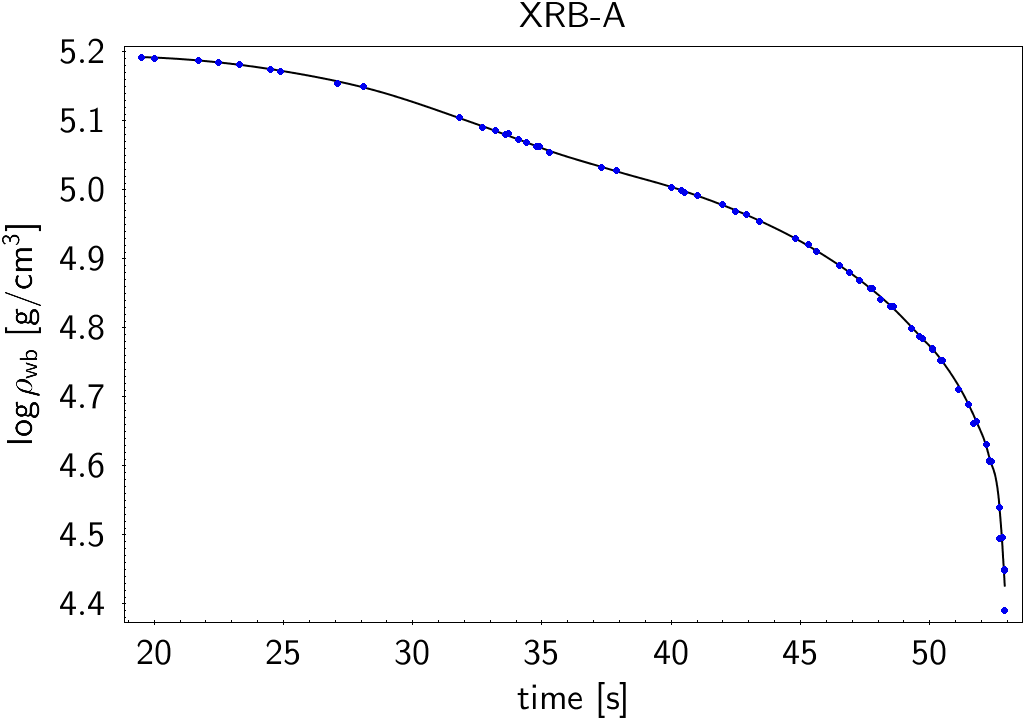} 
  \hfill
  \includegraphics[keepaspectratio=true,width=0.49\textwidth,clip=true,trim=0pt  0pt 0pt 40pt]{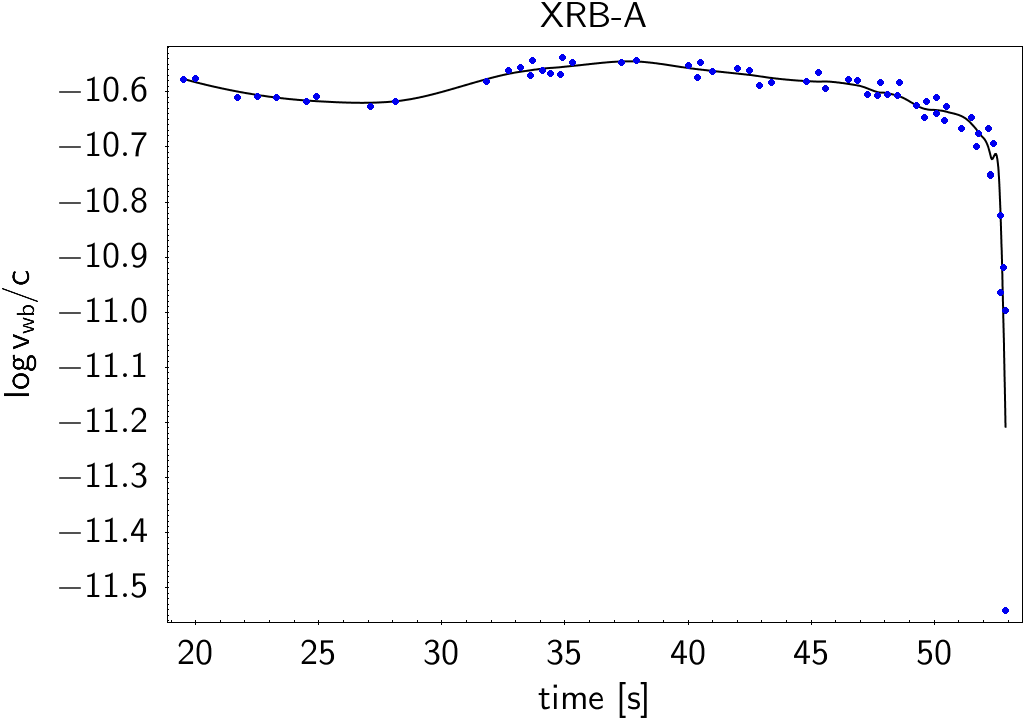}  
  \linebreak \linebreak
  \includegraphics[keepaspectratio=true,width=0.49\textwidth,clip=true,trim=0pt  0pt 0pt 40pt]{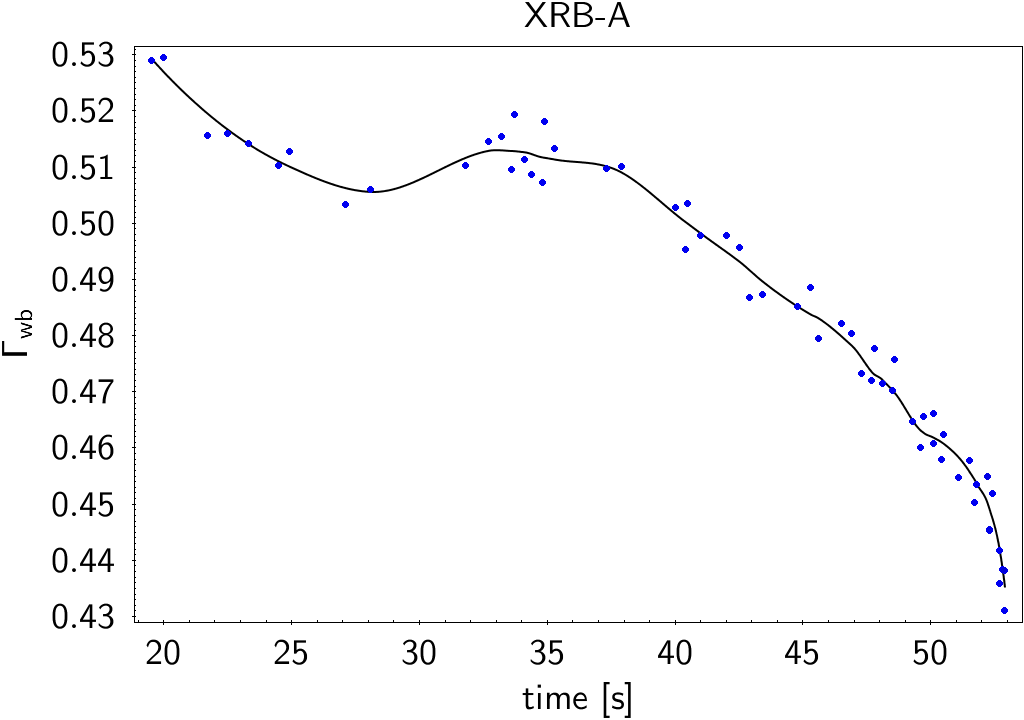}  
  \hfill
  \includegraphics[keepaspectratio=true,width=0.49\textwidth,clip=true,trim=0pt  0pt 0pt 40pt]{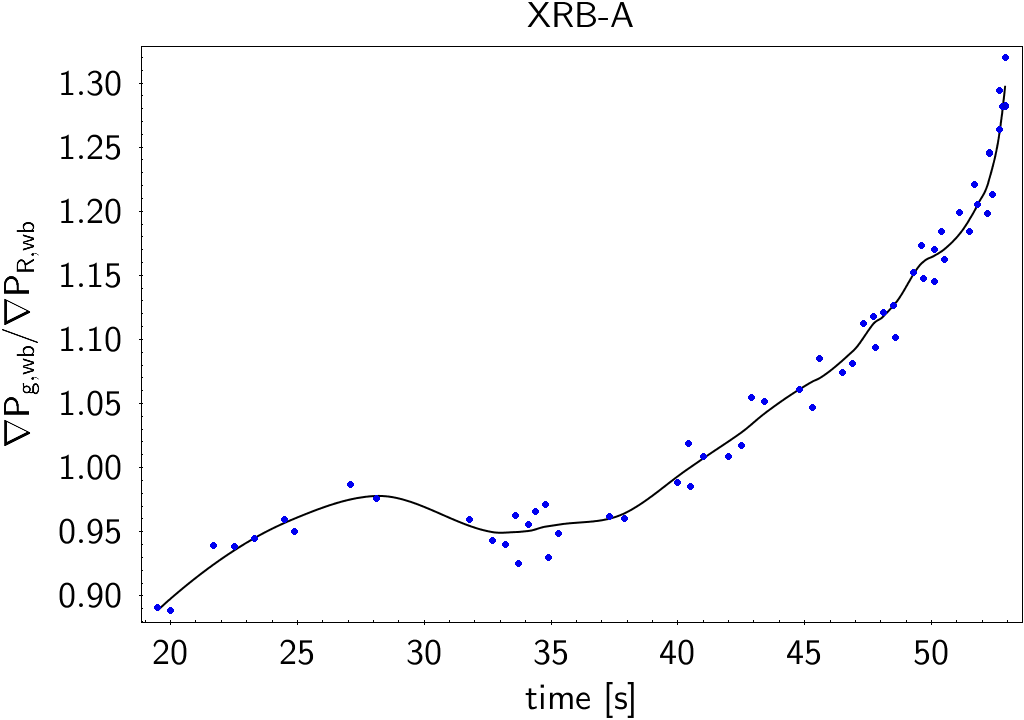}  
  \caption{ 
    Same as Fig. \ref{fig: XRB-Wind photosphere time smoothing}, for magnitudes evaluated at the wind base, 
    in reading order: radius, temperature, density, velocity luminosity ratio $\Gamma$, and gas-to-radiation pressure gradient ratio. 
    }
    \label{fig: XRB-Wind wind base time smoothing} 
\end{figure*}

  Finally, Figures \ref{fig: XRB-Wind critical time smoothing} and \ref{fig: XRB-Wind wind base time smoothing} show the time evolution of physical magnitudes at the critical point and wind base (XRB model matching point), respectively.
  The critical point remains much closer to the base, extending only a few times the NS radius, and receding at a somewhat uniform rate.
  Temperature, velocity and density at the critical point evolve in the opposite direction, slowly growing, with an overall smaller change of velocity around $v_\text{cr} \sim 10^{-3}c$, but with an abrupt drop right before the end of the wind.
  Luminosity stays very close below the local Eddington luminosity. 
  The effective optical depth of the critical point stays always higher than the required threshold of $10$, needed to consider the wind optically thick
  \cite[see][and references therein]{HSJ2020},
  but it seems to drop very quickly by the end of the wind. 
  This may hint that there could be a continuation of the wind with an optically thin regime, which our wind simulation code does not contemplate.
  However, the XRB-wind matching point moves quickly towards the outermost shells of the XRB hydrodynamic model where the boundary condition $P_g=0$ was set, making the appearance of a thin wind unlikely and, in practice, impossible to search further matches, even if the wind code was capable of dealing with an optically-thin wind.
  
  The ratio of gas pressure gradient to radiation pressure gradient at the wind base (last panel of Fig. \ref{fig: XRB-Wind wind base time smoothing}) shows values very close to unity. 
  This result rests in good agreement with the very ``generic'' boundary condition assumed for the characterization study in \cite{HSJ2020}.
  It does not come as a surprise, since the previously assumed boundary condition was inspired by (and is a necessary condition for) a situation in which wind velocity can be neglected and radiation pressure is not dominant, both of which correspond to a transition region between stellar wind and static envelope. 
  These turned to be good approximations to physical conditions at the wind base for all XRB hydrodynamic models studied, as can be seen for current model XRB-A (Fig. \ref{fig: XRB-Wind wind base time smoothing}, last three panels): wind velocity is smaller than $1\texttt{ cm/s}$, and radiative luminosity is about one half of the local Eddington limit needed to accelerate the wind; the remainder half of the driving force is provided by gas pressure gradient, due to the abrupt temperature and density changes.

\section{Conclusions and Discussion} \label{sect:discuss}

  We linked the stellar wind model developed in \cite{HSJ2020} to a series of XRB hydrodynamic models developed in \cite{Jose2010}.
  For this, we developed a technique that successfully matched different wind profiles to the boundary conditions given at different points during the evolution of each burst, with a quasi-stationary approach.
  This allowed us to construct a time evolution of wind profiles and to quantify the mass-loss of each isotope produced by nucleosynthesis during the burst.
  The overall ejected mass in XRB-A was about
  $\sim 6\E{19}\ \texttt{g}$.\footnote{
  Smaller values, in the range of $ \sim 2\E{17} - 7\E{18} \ \texttt{g}$, were found in lower resolution models. 
  Therefore, the resolution adopted could influence the specific ejected masses.
  }
  The average ejected mass per unit time represents 2.6\% of the accretion rate, with $0.1\%$ of the envelope mass ejected per burst and 90\% of the ejecta composed by $^{60}$Ni, $^{64}$Zn, $^{68}$Ge and $^{58}$Ni. Additionally, the amounts of $^{1}$H and $^{4}$He add up to less than $5\%$ of the ejected mass.
  
  The ejected material also contained a small fraction ($10^{-4}-10^{-5}$) of some light p-nuclei of interest. 
  The species with the highest amount 
  was $^{92}\text{Mo}$, with $\sim 1.5\E{16}\ \texttt{g}$ in XRB-A. 
  However, an estimation of their significance showed that an unreasonable number of XRB sources like the one analyzed (corresponding to a typical X-ray burst episode) was required to account for the Galactic abundances of each isotope. 
  When applied to the sequence of lower-resolution bursts \cite[i.e., Model 1 in][]{Jose2010}, with evolving metallicities and ignition conditions between them, our method showed varying quantities of these isotopes, but all of them smaller by at least an order of magnitude. 
  Therefore, we concluded that XRB sources are unlikely to constitute the sole explanation of their origin.
  
  Much as in \cite{HSJ2020}, the resulting wind profiles were all observed to transition from a gas pressure-driven regime at the inner parts, close to the wind base where opacity is low, into a full radiatively-driven wind in the outer parts. 
  Most notably, photospheric magnitudes showed the same correlations, namely $r_\text{ph} \sim T_\text{ph}^{-2} \sim \rho_\text{ph}^{-1}$, independently of the radius at which the aforementioned wind base condition was found. 
  These correlations are apparently independent from model parameters $(\dot M, \dot E)$ too, so they are expected to appear in every scenario.
  They seem to derive from the choice of boundary conditions at the photosphere and the fact that photospheric luminosity takes on values very close to Eddington luminosity $L_o =\frac{4\pi cGM}{\kappa_o}$, for every choice of parameters $(\dot M, \dot E)$.
  
  Another set of correlations found to hold in the present work involves parameters $(\dot M, \dot E)$ as well as photospheric magnitudes (see Eq. \ref{eq: photospheric correlations with parameters}). Since wind parameters are determined upon imposing conditions for all physical variables at the wind base, these correlations effectively link observable magnitudes to the physics of the innermost parts of the envelope, close to its interface with the NS core. 
  This could lead to a technique that allows to indirectly determine the radius of the NS, from photospheric magnitudes. 
  This possibility requires further study, but a general idea could be outlined as follows.
  For a fixed set of wind parameters ($\dot M$, $\dot E$, NS mass $M$, and gas composition $X_i$), one can determine the location of the wind base $r_\text{wb}$, understood now as the point where the wind profile 
  transitions from gas pressure-driven to radiation-driven, that is $\Nabla P_\text{gas} = \Nabla P_\text{rad}$. 
  Such wind base $r_\text{wb}$ remains very close ($\sim 20-60 \texttt{ m}$) to the actual NS radius (see Fig. \ref{fig: XRB-Wind wind base time smoothing}, first and last panels).
  Now, as $\dot M$, $\dot E$ evolve with the burst, they follow a path (see  Fig. \ref{fig: XRB-Wind WiMP}) that resembles remarkably those of constant $r_\text{wb}$ in parameter space from Fig. 3, panel 3 in \cite{HSJ2020}.
  The actual ($\dot M$, $\dot E$) path will not be exactly the same, since composition also evolves, but it would traverse a rather small range of values of $r_\text{wb}$ in an analogous parameter space plot calculated with an average value of $\mu(X_i)$ (which varies little, as stated previously when discussing the validity of quasi-stationary approach in Sect. \ref{sect: XRB-wind match analysis}). For high $\dot M$, and low $\dot E$, which in Model XRB-A corresponds to the earliest available times of the wind phase, this value of $r_\text{wb}$ would be closer to the actual NS radius. This could give us an estimation with an error as low as only a few tens of meters.

  Nevertheless, further study is required to determine whether these results are affected by the inclusion of more complex input physics, like relativistic wind models, or a more detailed treatment of radiative transfer.

\begin{acknowledgements}

This work has been partially supported by the Spanish MINECO grant PID2020-117252GB-I00, by the E.U. FEDER funds, and by the AGAUR/Generalitat de Catalunya grant
SGR-386/2021. 
This article benefited from discussions within the EU' H2020 project No. 101008324 ``ChETEC-INFRA''.
This work was also partly supported by the Spanish program Unidad de Excelencia Mar\'ia de Maeztu CEX2020-001058-M, financed by MCIN/AEI/10.13039/501100011033.
\end{acknowledgements}

\bibliographystyle{aa.bst} 
\bibliography{biblio.bib}

\end{document}